\documentclass[twocolumn,tighten]{aastex63}
\usepackage{color}

\usepackage{multirow}
\shortauthors{Sano, Yoshiike, Yamane et al. (2021)}
\shorttitle{ALMA CO Observations of the Mixed-Morphology SNR W49B}

\begin{document}
\title{ALMA CO Observations of the Mixed-Morphology Supernova Remnant W49B: Efficient Production of Recombining Plasma and Hadronic Gamma-rays via Shock--Cloud Interactions}

\author[0000-0003-2062-5692]{H. Sano}
\affiliation{National Astronomical Observatory of Japan, Mitaka, Tokyo 181-8588, Japan; hidetoshi.sano@nao.ac.jp}

\author[0000-0002-2458-7876]{S. Yoshiike}
\affiliation{Department of Physics, Nagoya University, Furo-cho, Chikusa-ku, Nagoya 464-8601, Japan}

\author[0000-0001-8296-7482]{Y. Yamane}
\affiliation{Department of Physics, Nagoya University, Furo-cho, Chikusa-ku, Nagoya 464-8601, Japan}

\author[0000-0001-6922-6583]{K. Hayashi}
\affiliation{Institute of Space and Astronautical Science (ISAS), Japan Aerospace Exploration Agency (JAXA), 3-1-1, Yoshinodai, Chuo-ku, Sagamihara, Kanagawa 252-5210, Japan}

\author[0000-0003-2735-3239]{R. Enokiya}
\affiliation{Department of Physics, Institute of Science and Technology, Keio University, 3-14-1 Hiyoshi, Kohoku-ku, Yokohama, Kanagawa 223-8522, Japan}

\author[0000-0002-2062-1600]{K. Tokuda}
\affiliation{National Astronomical Observatory of Japan, Mitaka, Tokyo 181-8588, Japan; hidetoshi.sano@nao.ac.jp}
\affiliation{Department of Physical Science, Graduate School of Science, Osaka Prefecture University, 1-1 Gakuen-cho, Naka-ku, Sakai 599-8531, Japan}

\author[0000-0002-1411-5410]{K. Tachihara}
\affiliation{Department of Physics, Nagoya University, Furo-cho, Chikusa-ku, Nagoya 464-8601, Japan}

\author[0000-0002-9516-1581]{G. Rowell}
\affiliation{School of Physical Sciences, The University of Adelaide, North Terrace, Adelaide, SA 5005, Australia}

\author[0000-0002-4990-9288]{M. D. Filipovi{\'c}}
\affiliation{Western Sydney University, Locked Bag 1797, Penrith South DC, NSW 1797, Australia}

\author[0000-0002-8966-9856]{Y. Fukui}
\affiliation{Department of Physics, Nagoya University, Furo-cho, Chikusa-ku, Nagoya 464-8601, Japan}

\begin{abstract}
We carried out new CO($J$ = 2--1) observations toward the mixed-morphology supernova remnant (SNR) W49B with the Atacama Large Millimeter/submillimeter Array (ALMA). We found that CO clouds at $\sim$10~km~s$^{-1}$ show a good spatial correspondence with synchrotron radio continuum as well as an X-ray deformed shell. The bulk mass of molecular clouds accounts for the western part of the shell, not for the eastern shell where near-infrared H$_2$ emission is detected. The molecular clouds at $\sim$10~km~s$^{-1}$ show higher kinetic temperature of $\sim$20--{60}~K, suggesting that modest shock-heating occurred. The expanding motion of the clouds with $\Delta V \sim$6~km~s$^{-1}$ was formed by strong winds from the progenitor system. We argue that the barrel-like structure of Fe rich ejecta was possibly formed not only by an asymmetric explosion, but also by interactions with dense molecular clouds. We also found a negative correlation between the CO intensity and the electron temperature of recombining plasma, implying that the origin of the high-temperature recombining plasma in W49B can be understood as the thermal conduction model. The total energy of accelerated cosmic-ray protons $W_\mathrm{p}$ is estimated to be $\sim$$2\times 10^{49}$~erg by adopting an averaged gas density of $\sim$$650\pm200$ cm$^{-3}$. The SNR age--$W_\mathrm{p}$ diagram indicates that W49B shows one of the highest in-situ values of $W_\mathrm{p}$ in the gamma-ray bright SNRs.
\end{abstract}
\keywords{Supernova remnants (1667); Interstellar medium (847); Cosmic ray sources (328); Gamma-ray sources (633); X-ray sources (1822)}

\section{INTRODUCTION}\label{sec:intro}
Mixed-morphology supernova remnants (SNRs) are characterized a center-filled thermal X-ray morphology with a synchrotron radio shell, accounting for more than 25\% of Galactic SNRs \citep{1998ApJ...503L.167R}. Most of the mixed-morphology SNRs are interacting with the dense interstellar medium (ISM) evidenced by radio-line emission such as CO, \ion{H}{1}, and/or 1720~MHz OH masers \citep[e.g.,][]{1998ApJ...505..286S,2003ApJ...585..319Y,2018ApJ...864..161K}. In addition, some of them are associated with GeV/TeV gamma-ray sources, which likely arise from interactions between accelerated cosmic-ray (CR) protons and dense clouds in the vicinity of the SNRs \citep[e.g.,][]{2008A&A...481..401A,2016PASJ...68S...5B}. Moreover, shock-propagation into the clumpy ISM and/or dense circumstellar matter (CSM) have a potential to explain their mixed-morphology and thermal X-ray radiation \citep[e.g.,][]{2012PASJ...64...24S,2017ApJ...846...77S,2019ApJ...875...81Z}. Therefore, the shock-interacting ISM plays an important role in understanding their morphology, plasma conditions, and cosmic-ray acceleration \citep[see also reviews by][]{2012A&ARv..20...49V,2020AN....341..150Y,2021arXiv210600708S}. To unveil the physical processes and high-energy phenomena in the mixed-morphology SNRs, detailed comparative studies among the radio-line emission, X-rays, and gamma-rays are needed.

\begin{figure}[]
\begin{center}
\vspace*{0.2cm}
\includegraphics[width=\linewidth,clip]{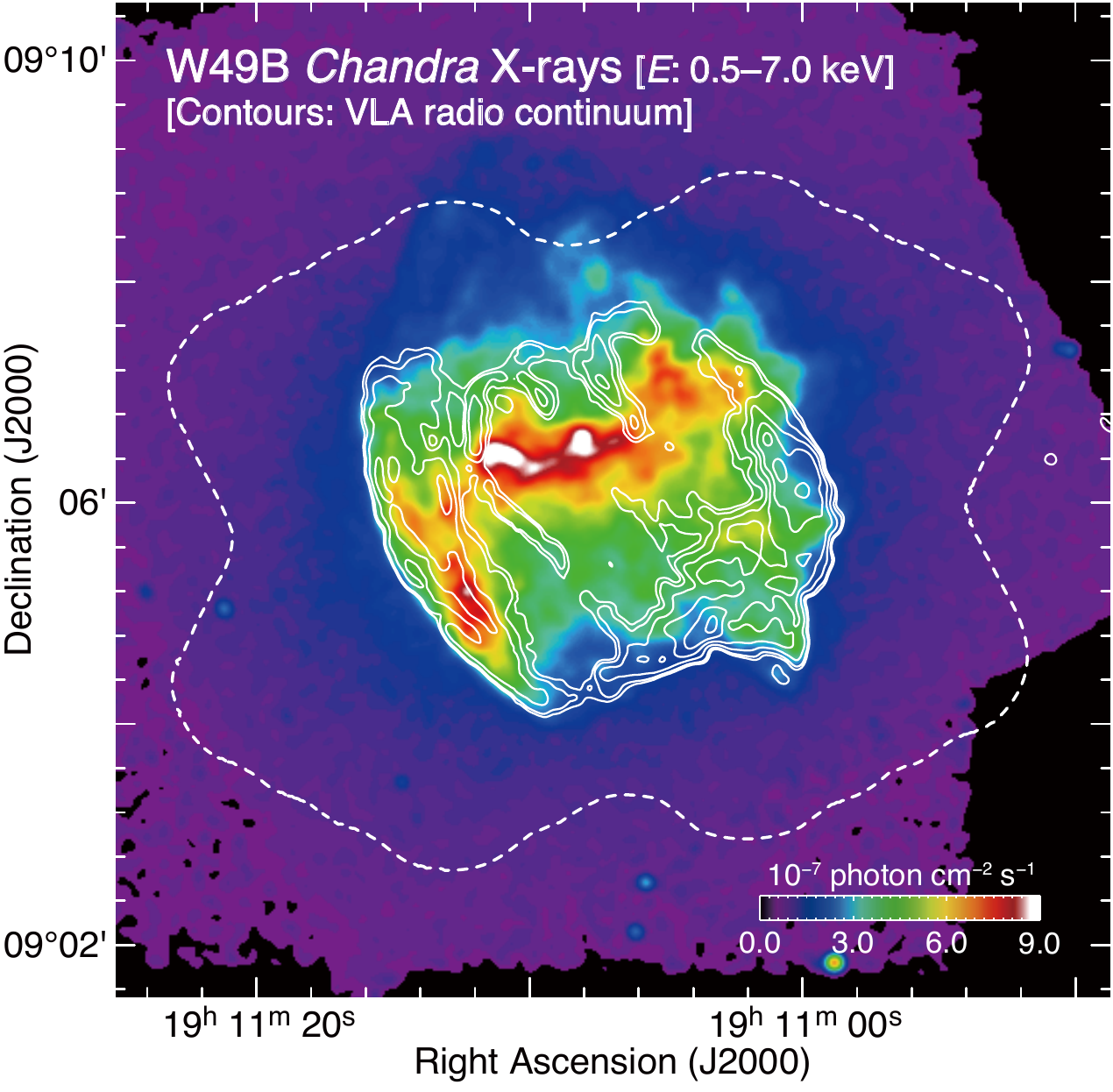}
\caption{Intensity map of {\it{Chandra}} broad-band X-rays \citep[$E$: 0.5--7.0~keV, e.g.,][]{2013ApJ...764...50L} superposed on the VLA radio continuum contours at 20~cm obtained from MAGPIS \citep{2006AJ....131.2525H}. The contour levels are 5.0, 7.6, 15.4, 28.4, 46.6, and 70.0 mJy beam$^{-1}$. The region enclosed by dashed rectangle corresponds to the observed region with ALMA ACA.}
\label{fig1}
\end{center}
\vspace*{0.5cm}
\end{figure}

W49B (also known as G43.3$-$0.2) is a well-studied Galactic mixed-morphology SNR with the bright radio-continuum shell and thermal-dominated center-filled X-rays as shown in Figure \ref{fig1}. The SNR is thought to be lying on the far-side of the Galaxy from us \citep[e.g.,][]{1978A&A....67..355L,2001ApJ...550..799B}. The small apparent diameter of $\sim$$3'$--$5'$ is consistent with the larger distance of $\sim$7.5--11.3~kpc \citep[][]{2014ApJ...793...95Z,2018AJ....155..204R,2020AJ....160..263L} and its young age \citep[5--6~kyr,][]{2018AA...615A.150Z}. W49B is also thought to be an efficient accelerator of CR protons owing to its bright GeV/TeV gamma-rays with the pion-decay bump \citep{2018A&A...612A...5H}. The total energy of CR protons was derived to be $\sim$10$^{49}$--10$^{51}$~erg, assuming the targeted gas density of 10--1000~cm$^{-3}$ \citep{2010ApJ...722.1303A}.

The X-ray radiation of W49B is characterized by three properties: the most luminous in Fe K-shell line emission \citep{2014ApJ...785L..27Y}, non-thermal Bremsstrahlung \citep{2018ApJ...866L..26T}, and recombining (overionized) plasma where the ionization temperature goes even higher than the electron temperature \citep{2009ApJ...706L..71O,2010A&A...514L...2M,2013ApJ...777..145L,2018AA...615A.150Z,2018ApJ...868L..35Y,2020ApJ...893...90S,2020ApJ...903..108H,2020ApJ...904..175S}. The elongated structure of Fe-rich ejecta is believed to be related to a bipolar/jet-driven Type Ib/Ic explosion and/or interactions between the shock and a surrounding interstellar cloud \citep{2007ApJ...654..938K,2013ApJ...764...50L,2017MNRAS.468..140B}. On the other hand, recent X-ray studies on metal abundances favor Type Ia models \citep{2018AA...615A.150Z,2020ApJ...904..175S}. Additionally, \cite{2020ApJ...893...90S} conclude that the SN type is unclear, with neither core-collapse or Ia models perfectly reproducing their best-fit abundances. The origin of recombining plasma---thermal conduction with cold-dense clouds and/or adiabatic cooling---is still being debated \citep[e.g.,][]{2018ApJ...868L..35Y,2020ApJ...893...90S,2020ApJ...903..108H}. If we detect decreasing the electron temperature toward the shocked clouds, we can confirm the thermal conduction scenario as a formation mechanism of the recombination plasma \citep[e.g.,][]{2017ApJ...851...73M,2018PASJ...70...35O,2020ApJ...890...62O}.

Although W49B is thought to be interacting with interstellar clouds, it is a perplexing question which clouds are physically associated. \cite{2007ApJ...654..938K} discovered 2.12~$\mu$m shocked H$_2$ emission toward the eastern and southwestern shells by near-infrared photometric observations. The total mass of shocked H$_2$ is estimated to 14--550~$M_{\sun}$. Subsequent radio observations using CO line emission revealed three molecular clouds at velocities of $\sim$10, $\sim$40, and $\sim$60~km~s$^{-1}$, which are possibly associated with the SNR \citep[e.g.,][]{2014ApJ...793...95Z,2016ApJ...816....1K,2020AJ....160..263L}. \cite{2014ApJ...793...95Z} argued that the molecular cloud at $\sim$40~km~s$^{-1}$ is interacting with W49B because of its wind-bubble like morphology. On the other hand, \cite{2016ApJ...816....1K} found a line-broadening of $^{12}$CO profile in a molecular cloud at $\sim$10~km~s$^{-1}$ located toward the western shell of W49B, and hence the authors claimed that the cloud at 10~km~s$^{-1}$ is interacting with the SNR. The velocity is roughly consistent with the \ion{H}{1} absorption studies \citep{2001ApJ...550..799B,2018AJ....155..204R}. Most recently, \cite{2020AJ....160..263L} performed near-infrared spectroscopy of shocked H$_2$ emission toward four strips on W49B. The authors found that a central velocity of shocked H$_2$ is $\sim$64~km~s$^{-1}$ and then concluded that the molecular cloud at $\sim$60~km~s$^{-1}$ located toward the center and the southwest shell of W49B is likely associated with the SNR. In either case, detailed spatial and kinematic studies as well as deriving cloud properties (e.g., mass, density, kinetic temperature) have not been performed due to the modest sensitivity and angular resolution of CO datasets up to $\sim$$20''$, corresponding to a spatial resolution of $\sim$1~pc at the distance of 10~kpc.

In the present paper, we report on results of new millimeter wavelength observations using CO($J$ = 2--1) line emission with the Atacama Compact Array (ACA, also known as Morita Array) which is a part of the Atacama Large Millimeter/submillimeter Array (ALMA). The unprecedented sensitivity and high-angular resolution of $\sim$$7''$ ($\sim$0.3~pc at the distance of 10~kpc) of the ALMA CO data enable us to identify the interacting molecular cloud and its physical relation to the high-energy phenomena in W49B. Section \ref{sec:obs} describes the observational datasets and reductions. Section \ref{sec:results} comprises five subsections: Sections \ref{subsec:overview}--\ref{subsec:pv} present overview distributions of X-ray, radio continuum, and CO; Section \ref{subsec:ratio}--\ref{subsec:lvg} show physical conditions of molecular clouds. Discussion and conclusions are given in Sections \ref{sec:discussion} and \ref{sec:conclusions}, respectively.

\section{OBSERVATIONS AND DATA REDUCTIONS}\label{sec:obs}
\subsection{CO}\label{subsec:co}
Observations of $^{12}$CO($J$ = 2--1) and $^{13}$CO($J$ = 2--1) line emission were conducted using ALMA ACA Band 6 (211--275 GHz) as a Cycle 6 project (proposal no. 2018.1.01780.S). We used the mosaic observation mode with 10--12 antennas of 7-m array and four antennas of 12-m total power (TP) array. The observed areas were $5\farcm1 \times 2\farcm7$ rectangular regions centered at ($\alpha_\mathrm{J2000}$, $\delta_\mathrm{J2000}$) $=$ ($19^\mathrm{h}11^\mathrm{m}09\fs00$, $+9\arcdeg06\arcmin24\farcs8$) and ($19^\mathrm{h}11^\mathrm{m}07\fs44$, $+9\arcdeg05\arcmin56\farcs2$). The actual observed area is shown in Figure \ref{fig1}. The combined baseline length of 7-m array data is from 8.85 to 48.95 m, corresponding to $u$--$v$ distances from 6.8 to 37.6 $k\lambda$ at 230.538 GHz. Two quasars, J1924$-$2914 and J1751$+$0939, were observed as bandpass and flux calibrators. We also observed four quasars, J1907$+$0127, J1922$+$1530, J1938$+$0448, and J1851$+$0035, as phase calibrators. We performed data reduction using the Common Astronomy Software Application \citep[CASA,][]{2007ASPC..376..127M} package version 5.5.0. We utilized ``tclean'' task with multi-scale deconvolver and natural weighting. The emission mask was also selected using the auto-multithresh procedure \citep{2020PASP..132b4505K}. We combined the cleaned 7-m array data and the calibrated TP array data using ``feather'' task to recover the missing flux and diffuse emission. The beam size of feathered data is $8\farcs23 \times 4\farcs77$ with a position angle of $-75.31\degr$ for the $^{12}$CO($J$ = 2--1) data, and $8\farcs28 \times 5\farcs04$ with a position angle of $-79.37\degr$ for the $^{13}$CO($J$ = 2--1) data. The typical noise fluctuations are $\sim$0.065 K for the $^{12}$CO($J$ = 2--1) data and $\sim$0.055 K for the $^{13}$CO($J$ = 2--1) data at the velocity resolution of 0.4 km~s$^{-1}$.

We also used archival datasets of $^{12}$CO($J$ = 1--0) and $^{12}$CO($J$ = 3--2) line emission for estimating physical properties of molecular clouds. The $^{12}$CO($J$ = 1--0) data are from the FOREST Unbiased Galactic Plane Imaging survey with the Nobeyama 45~m telescope \citep[FUGIN,][]{2017PASJ...69...78U}, and the $^{12}$CO($J$ = 3--2) data are from the CO High-Resolution Survey \citep[COHRS,][]{2013ApJS..209....8D} obtained with the James Clerk Maxwell Telescope (JCMT). The angular resolution is $\sim$$20''$ for the $^{12}$CO($J$ = 1--0) data and $\sim$16\farcs6 for the $^{12}$CO($J$ = 3--2) data. The velocity resolutions of $^{12}$CO($J$ = 1--0) and $^{12}$CO($J$ = 3--2) data are 1.3 and 1.0 km s$^{-1}$, respectively. To improve the signal-to-noise ratio of the $^{12}$CO($J$ = 3--2) data, we combined four spatial pixels and a rebind pixel size is to be $12''$. The typical noise fluctuations are $\sim$1.4 K for the $^{12}$CO($J$ = 1--0) data and $\sim$0.18 K for the $^{12}$CO($J$ = 3--2) data at each velocity resolution.

\subsection{Radio Continuum}\label{subsec:rc}
The radio continuum data at 20 cm wavelength are from the Multi-Array Galactic Plane Imaging Survey \citep[MAGPIS,][]{2006AJ....131.2525H} obtained with the Very Large Array (VLA) and the Effelsberg 100-m telescope. The angular resolution is $\sim$$6''$, which is compatible for the ALMA ACA resolution. The typical noise fluctuations are $\sim$1--2 mJy. 

\subsection{X-rays}\label{subsec:xrays}
We utilized archival X-ray data obtained by {\it{Chandra}} (observation IDs are 117, 13440, and 13441), which have been published in numerous papers \citep[e.g.,][]{2005ApJ...631..935K,2009ApJ...691..875L,2009ApJ...706L.106L,2011ApJ...732..114L,2013ApJ...777..145L,2013ApJ...764...50L,2009ApJ...692..894Y,2016ApJ...821...20K,2018AA...615A.150Z}. The X-ray datasets were taken with the Advanced CCD Imaging Spectrometer S-array (ACIS-S3). We used Chandra Interactive Analysis of Observations \citep[CIAO,][]{2006SPIE.6270E..1VF} software version 4.12 with CALDB 4.9.1 \citep[][]{2007ChNew..14...33G} for data reduction and imaging. After reprocessing for all datasets using the ``chandra\_repro'' procedure, we created an energy-filtered, exposure-corrected image using the ``fluximage'' procedure in the energy bands of 0.5--7.0~keV (broad-band, see Figure \ref{fig1}), 0.5--1.2 keV (soft-band), 1.2--2.0 keV (medium band), 2.0--7.0 keV (hard-band), and 4.2--5.5 keV (continuum band). Because the soft- and medium-band images are heavily affected by interstellar absorption \cite[e.g.,][]{2018AA...615A.150Z}, in this paper we focus on the X-ray images at energies greater than 2.0 keV. We also created an exposure-corrected, continuum-subtracted image of Fe He$\alpha$ line emission (6.4--6.9 keV) following the method presented by \cite{2013ApJ...764...50L}. The typical angular resolution of {\it{Chandra}} images is $\sim$0\farcs5.

\begin{figure*}[]
\begin{center}
\includegraphics[width=\linewidth,clip]{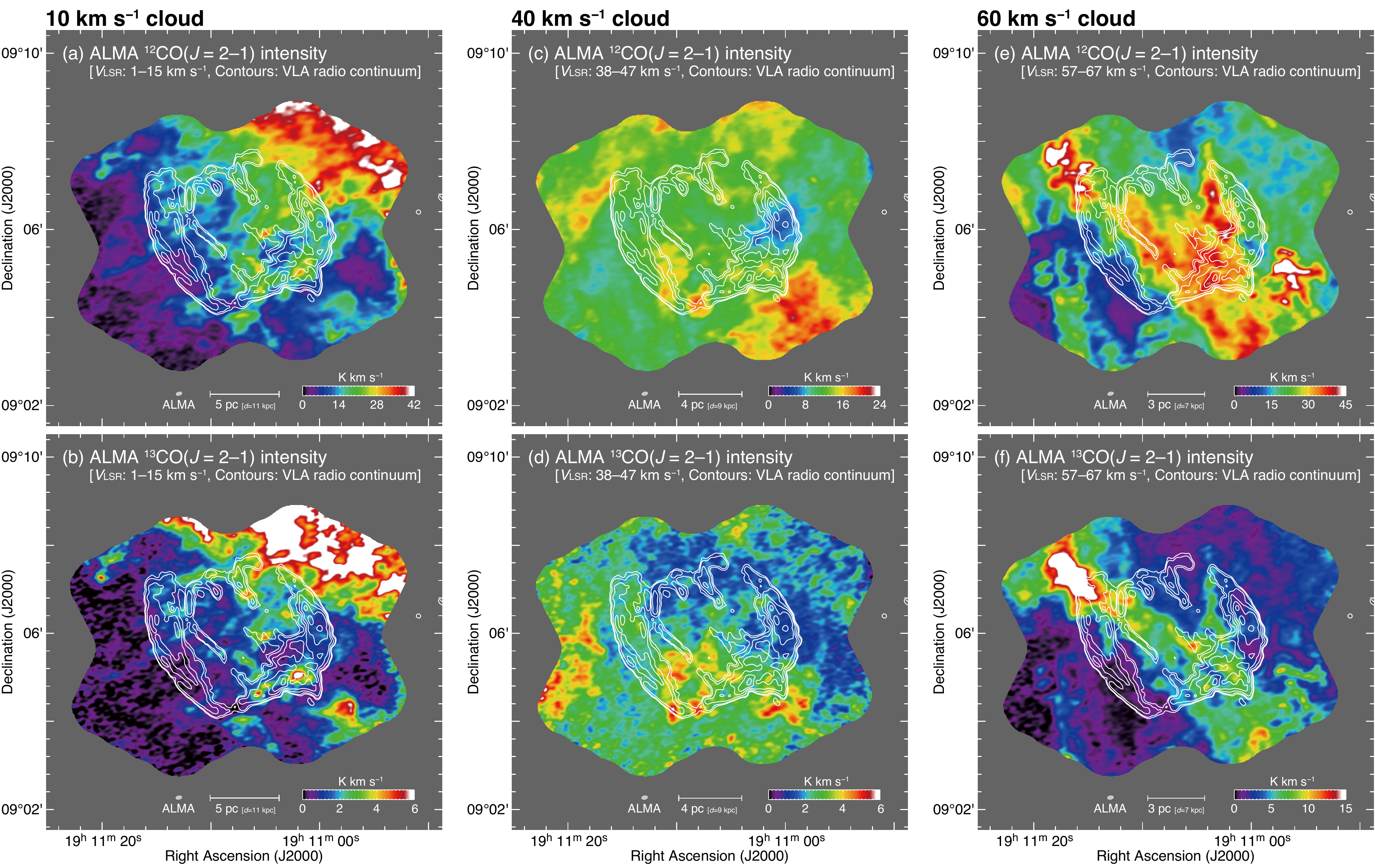}
\caption{Integrated intensity maps of ALMA ACA $^{12}$CO($J$ = 2--1) ({\it{upper panels}}) and $^{13}$CO($J$ = 2--1) ({\it{lower panels}}) for (a, b) the 10~km~s$^{-1}$ cloud, (c, d) the 40~km~s$^{-1}$ cloud, and (e, f) the 60~km~s$^{-1}$ cloud. The integration velocity range is from 1 to 15 km~s$^{-1}$ for the 10~km~s$^{-1}$ cloud; from 38 to 47 km~s$^{-1}$ for the 40~km~s$^{-1}$ cloud; and is from 57 to 67 km~s$^{-1}$ for the 60~km~s$^{-1}$ cloud. Superposed contours are the same as shown in Figure \ref{fig1}.}
\label{fig2}
\end{center}
\end{figure*}%

\section{RESULTS}\label{sec:results}
\subsection{Overview of X-ray, Radio Continuum, and CO Distributions}\label{subsec:overview}
Figure \ref{fig1} shows the {\it{Chandra}} broad-band X-ray image of W49B superposed on the VLA radio continuum contours at 20 cm wavelength. As presented in previous studies, a barrel-shaped radio-continuum shell with several co-axis filaments and the center-filled X-rays are seen \citep[e.g.,][]{2007ApJ...654..938K,2009ApJ...691..875L,2011ApJ...732..114L}. The X-ray elongated feature brighter than $\sim5 \times 10^{-7}$ photon cm$^{-2}$ s$^{-1}$ is roughly consistent with the spatial distribution of Fe He$\alpha$ emission (see Figure 3 in \citeauthor{2013ApJ...764...50L}~\citeyear{2013ApJ...764...50L}). We note that overall distributions of X-rays and radio continuum are quite different between the northeastern and southwestern halves: the shell boundary of the northeastern half roughly coincides with each other, whereas the southwestern shell of X-rays is significantly deformed compared to that of radio continuum. In particular, the X-rays are dim around positions at ($\alpha_\mathrm{J2000}$, $\delta_\mathrm{J2000}$) $\sim$ ($19^\mathrm{h}11^\mathrm{m}00\fs0$, $+09\arcdeg06\arcmin00''$), ($19^\mathrm{h}11^\mathrm{m}03\fs0$, $+09\arcdeg05\arcmin00''$), and ($19^\mathrm{h}11^\mathrm{m}08\fs0$, $+09\arcdeg06\arcmin06''$): the first two correspond to the bright peaks of radio-continuum and the other is partially surrounded by radio filaments. This trend is also seen in the 4.2--5.5 keV band image which is mostly free from the interstellar absorption as well as line-emission.

\begin{figure*}[]
\begin{center}
\includegraphics[width=\linewidth,clip]{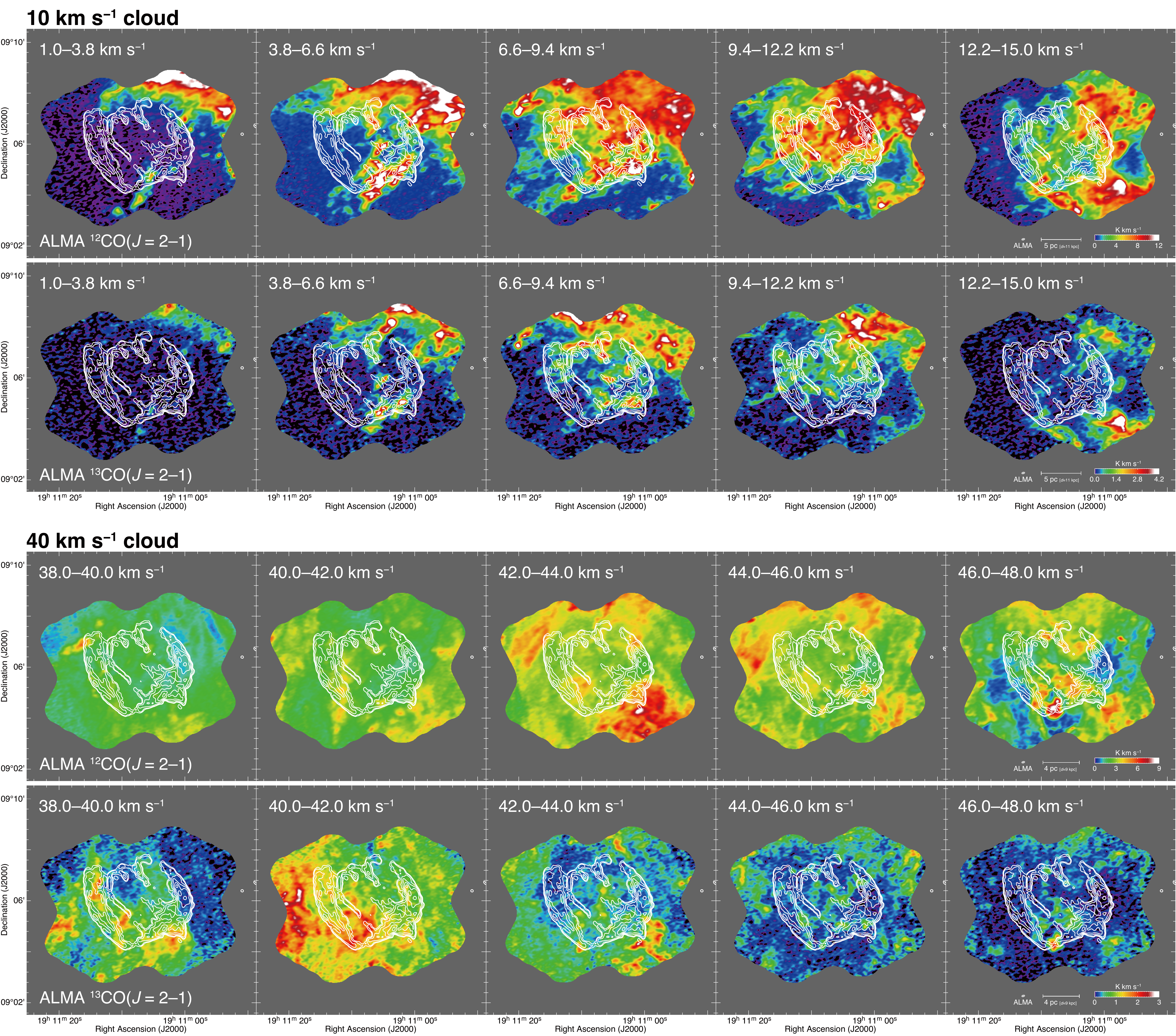}
\caption{Velocity channel maps of ALMA ACA $^{12}$CO($J$ = 2--1) ({\it{upper panels}}) and $^{13}$CO($J$ = 2--1) ({\it{lower panels}}) for each cloud. Each panel shows CO integrated intensity distribution integrated over the velocity range from 1 to 15 km s$^{-1}$ every 2.8 km s$^{-1}$ for the 10~km~s$^{-1}$ cloud; from 38 to 48 km s$^{-1}$ every 2 km s$^{-1}$ for the 40~km~s$^{-1}$ cloud; and 57 to 67 km s$^{-1}$ every 2 km s$^{-1}$ for the 60~km~s$^{-1}$ cloud. Superposed contours are the same as shown in Figure \ref{fig1}.}
\label{fig3}
\end{center}
\end{figure*}%

\setcounter{figure}{2}
\begin{figure*}[]
\begin{center}
\includegraphics[width=\linewidth,clip]{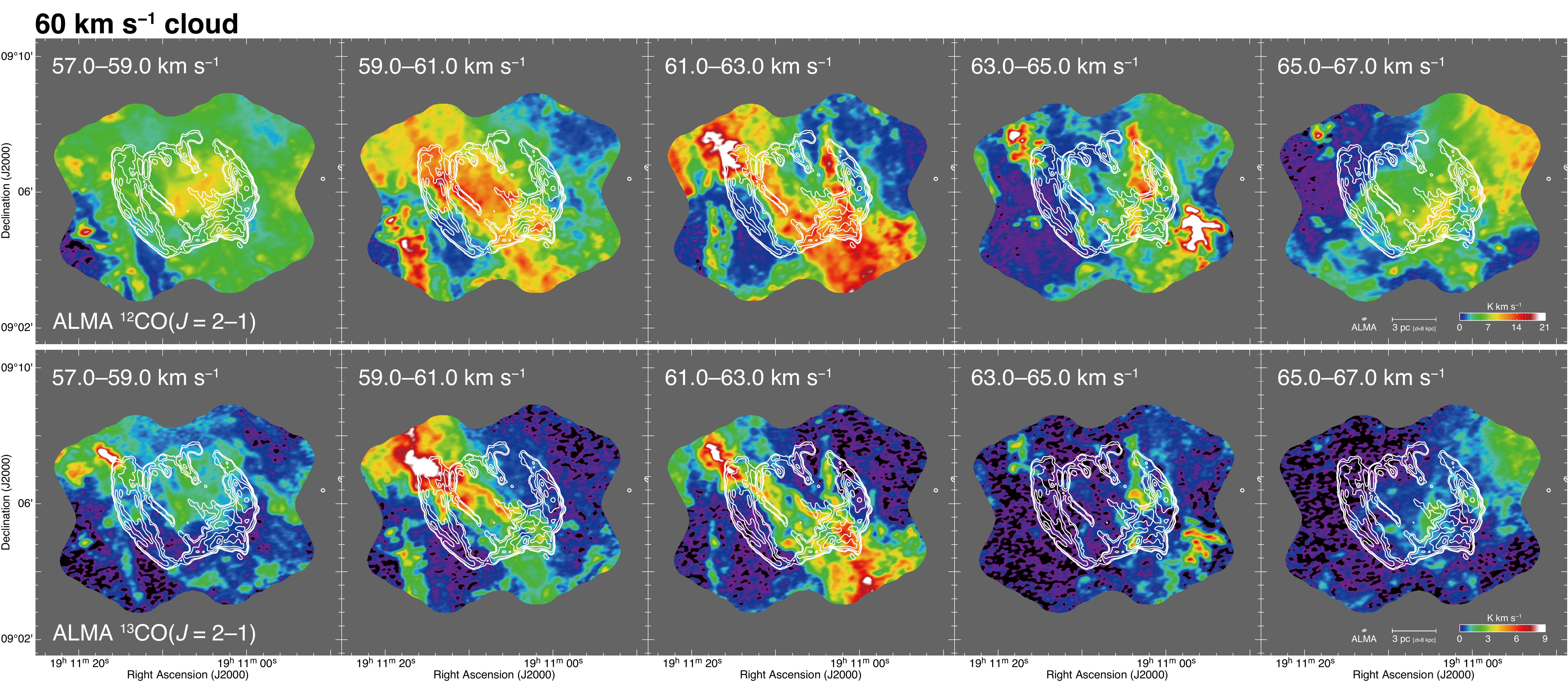}
\caption{{\it{Continued.}}}
\end{center}
\end{figure*}%

Figure \ref{fig2} shows integrated intensity maps of $^{12}$CO($J$ = 2--1) and $^{13}$CO($J$ = 2--1) for three velocity ranges of 1--15 km s$^{-1}$ (hereafter ``10 km s$^{-1}$ cloud''), 38--47 km s$^{-1}$ (hereafter ``40 km s$^{-1}$ cloud''), and 57--67 km s$^{-1}$ (hereafter ``60 km s$^{-1}$ cloud'') as previously mentioned in several papers \citep[e.g.,][]{2014ApJ...793...95Z,2016ApJ...816....1K,2020AJ....160..263L}. The kinematic distance of molecular cloud is $\sim$11 kpc for the 10 km s$^{-1}$ cloud; $\sim$9 kpc for the 40 km s$^{-1}$ cloud \footnote{Although \cite{2014ApJ...793...95Z} suggested the distance of 40 km s$^{-1}$ cloud to be $\sim$10~kpc using a Galactic rotation curve model with $R_0 = 8.5$~kpc and $\Theta_0 = 220$~km s$^{-1}$ \citep{1986MNRAS.221.1023K}, we adopt its distance to be $\sim$9~kpc using the latest Galactic parameters of $R_0 = 7.92$ kpc and $\Theta_0 = 227$ km s$^{-1}$ \citep{2020PASJ...72...50V}. Here $R_0$ is the distance from the Sun to the Galactic center and $\Theta_0$ is the rotation velocity of the local standard of rest. We use the latter values throughout the paper.}; and $\sim$7 kpc for the 60 km s$^{-1}$ cloud \citep{2020arXiv201111916S}. Note that there are no other CO clouds within the velocity range from $-15.0$ to 92.6 km s$^{-1}$, and hence we focus on the three molecular clouds in the present paper. 

In the 10 km s$^{-1}$ cloud (Figures \ref{fig2}a and \ref{fig2}b), there is an intensity gradient increasing from southeast to northwest. The radio continuum shows fairly good spatial correspondence with molecular clouds in the northern inward protrusion and along the north, northwest, and southwest rims. On the other hand, both the $^{12}$CO and $^{13}$CO emission lines are faint in the southeastern shell where shocked H$_2$ emission is strongly detected \citep[e.g.,][]{2007ApJ...654..938K}. Dense clouds traced by $^{13}$CO emission are located not only outside the shell boundary, but also inside the radio continuum shell. 

In the 40 km s$^{-1}$ cloud (Figures \ref{fig2}c and \ref{fig2}d), the $^{12}$CO emission has a relatively uniform distribution rather than that of the 10 km s$^{-1}$ cloud, but the $^{13}$CO emission shows an intensity gradient increasing from northwest to southeast. We note that the radio brightest shell in west shows lack of both the $^{12}$CO and $^{13}$CO emission lines. The bright $^{12}$CO emission and $^{13}$CO clumps are located both toward the southwest of the SNR as well as inside the southeastern shell. The southwestern CO clumps seem to be along the sharp edge of radio shell, whereas the CO clumps inside the SNR show no significant spatial correlations with the radio shell morphology. At this spatial coverage, we could not find the bubble-like CO structure toward W49B as mentioned by \cite{2014ApJ...793...95Z}.

In the 60 km s$^{-1}$ cloud (Figures \ref{fig2}e and \ref{fig2}f), there are dense clouds across the SNR from northeast to southwest with two bright CO peaks at ($\alpha_\mathrm{J2000}$, $\delta_\mathrm{J2000}$) $\sim$ ($19^\mathrm{h}11^\mathrm{m}17\fs0$, $+09\arcdeg07\arcmin30''$) and ($19^\mathrm{h}10^\mathrm{m}57\fs5$, $+09\arcdeg05\arcmin07''$). The former contains two \ion{H}{2} regions cataloged by \cite{2009A&A...501..539U}, whereas the latter does not have any cataloged objects. The diffuse CO emission inside the SNR appears to be spatially anti-correlated with the radio continuum contours. 

To derive masses of three molecular clouds, we used the following equations:
\begin{eqnarray}
M = m_{\mathrm{H}} \mu \Omega D^2 \sum_{i} N_i(\mathrm{H}_2),\\
N(\mathrm{H}_2) = X \cdot W(\mathrm{CO}),
\label{eq1}
\end{eqnarray}
where $m_\mathrm{H}$ is the mass of hydrogen, $\mu = 2.8$ is the mean molecular weight, $\Omega$ is the solid angle of each pixel, $D$ is the distance to W49B, $N_i(\mathrm{H}_2)$ is the column density of molecular hydrogen for each pixel, $X$ is CO-to-H$_2$ conversion factor of $2 \times 10^{20}$~cm$^{-2}$ (K~km~s$^{-1})^{-1}$ \citep{1993ApJ...416..587B}, and $W$(CO) is the velocity integrated intensity of $^{12}$CO($J$ = 1--0) emission line obtained from the FUGIN data \citep{2017PASJ...69...78U}. We estimated the mass of molecular cloud inside the radio shell to be $\sim$$4.1 \times 10^4$ $M_{\sun}$ for the 60 km s$^{-1}$ cloud and $\sim$$2.7 \times 10^4$ $M_{\sun}$ for the other two clouds, where we adopted the shell radius of 2\farcm5 centered at ($\alpha_\mathrm{J2000}$, $\delta_\mathrm{J2000}$) $=$ ($19^\mathrm{h}11^\mathrm{m}07\fs34$, $+09\arcdeg06\arcmin01\farcs1$). These values are roughly consistent with previously derived cloud masses using the $^{13}$CO($J$ = 1--0) emission and the $^{13}$CO-to-H$_2$ conversion factor \citep{2018A&A...612A...5H}.

\subsection{Detailed Spatial Comparison with the Radio Continuum Shell}\label{subsec:comprc}
Figure \ref{fig3} shows velocity channel maps for each molecular cloud superposed on the radio continuum contours. We find that dense $^{13}$CO clumps at a velocity rang from 3.8 to 9.4 km s$^{-1}$ are nicely along not only with the northern and southern shell, but also with radio filaments inside the SNR at positions of ($\alpha_\mathrm{J2000}$, $\delta_\mathrm{J2000}$) $\sim$ ($19^\mathrm{h}11^\mathrm{m}06\fs4$, $+09\arcdeg07\arcmin03''$) and ($19^\mathrm{h}11^\mathrm{m}05\fs5$, $+09\arcdeg05\arcmin56''$). We also find that both the $^{12}$CO and $^{13}$CO clouds at the velocity range of 9.4--12.2 km s$^{-1}$ show global anti-correlation with the radio shell. In addition, $^{12}$CO emission at the velocity range of 12.2--15.0 km s$^{-1}$ shows a good spatial correspondence with the western shell especially for the sharp edge of the southwestern rim. This velocity range is consistent with that inferred from Kilpatrick et al. (2016). The southwestern CO clumps at 42.0--44.0 km s$^{-1}$ seem to be along the sharp edge of radio shell especially prominent in the $^{13}$CO line emission. Moreover, the $^{13}$CO clouds at 38.0--40.0 km s$^{-1}$ shows the good spatial correspondence with the southeastern half of the radio continuum shell. Furthermore, the 63--65 km s$^{-1}$ $^{12}$CO map shows a lack of CO emission along the eastern shell and shows bright CO emission in the gaps between the western and northern radio contours. Although the other CO clouds also appear to be overlapped with the radio continuum shell and filaments, their spatial correspondence is not clear.

\begin{figure*}[]
\begin{center}
\includegraphics[width=\linewidth,clip]{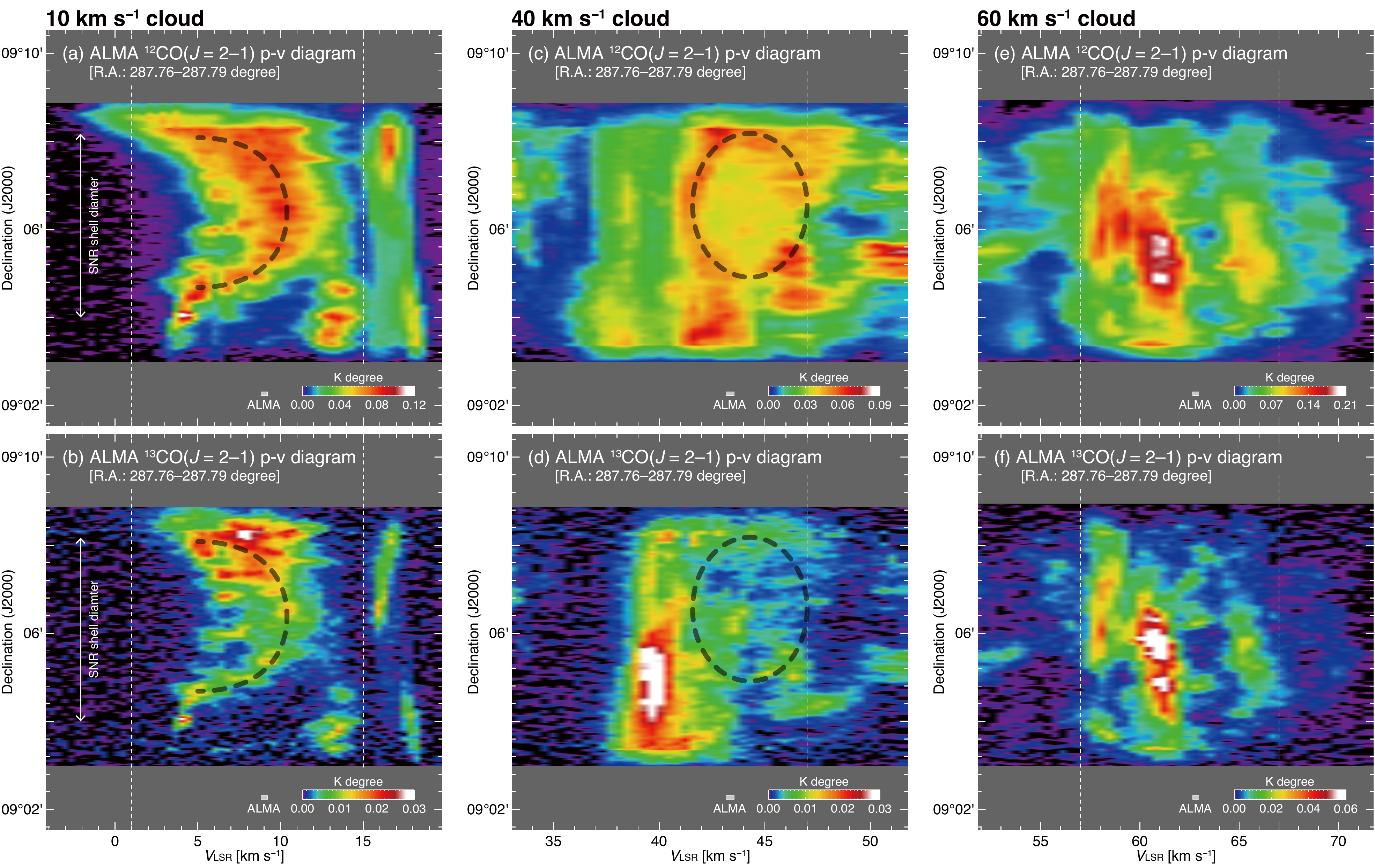}
\caption{Position-velocity diagram of ALMA ACA $^{12}$CO($J$ = 2--1) ({\it{upper panels}}) and $^{13}$CO($J$ = 2--1) ({\it{lower panels}}) for (a, b) the 10~km~s$^{-1}$ cloud, (c, d) the 40~km~s$^{-1}$ cloud, and (e, f) the 60~km~s$^{-1}$ cloud. The integration range of Right Ascension is from 283\fdg76 to 287\fdg79. Dashed curves and circles delineate expanding gas motion (see the text). Vertical dashed lines indicate integration velocity ranges for each cloud.}
\label{fig4}
\end{center}
\end{figure*}%

\subsection{Position--Velocity Diagrams}\label{subsec:pv}
Figure \ref{fig4} shows position--velocity diagrams for the three molecular clouds. The velocity distributions of $^{12}$CO and $^{13}$CO emission are similar to each other except for a velocity at $\sim$40 km s$^{-1}$, suggesting that the $^{12}$CO emission at $\sim$40 km s$^{-1}$ is subject to self-absorption due to optically thick component. In fact, the $^{12}$CO emission at $\sim$40 km s$^{-1}$ shows dip-like feature, whereas the $^{13}$CO spectrum at the same velocity has strong line emission. We also find incomplete and complete cavity-like structures in the 10 and 40 km s$^{-1}$ clouds, respectively (see dashed curves in Figures \ref{fig4}a--\ref{fig4}d). The incomplete cavity (or arc-like distribution) in the 10 km s$^{-1}$ cloud lies from 5 km s$^{-1}$ to 11 km s$^{-1}$ with a velocity dispersion of a few km s$^{-1}$. On the other hand, the complete cavity in the 40 km s$^{-1}$ cloud is clearly seen especially for $^{12}$CO emission, whose velocity range is from 41 km s$^{-1}$ to 47 km s$^{-1}$. Note that spatial extents of these cavities are roughly consistent with the diameter of the radio continuum shell. By contrast, there is no clear evidence for such cavity-like structure in the position--velocity diagram of the 60 km s$^{-1}$ cloud (see Figures \ref{fig4}e and \ref{fig4}f).

\subsection{Intensity Ratio Maps}\label{subsec:ratio}
Figure \ref{fig5} shows intensity ratio maps of $^{12}$CO($J$ = 3--2) / $^{13}$CO($J$ = 2--1) (hereafter $R_\mathrm{12CO32/13CO21}$) toward the three molecular clouds. A higher value of $R_\mathrm{12CO32/13CO21}$ tends to be observed in diffuse warm gas thermalized by supernova shocks and/or stellar radiation assuming that the abundance ratio of $^{12}$CO/$^{13}$CO is constant within a molecular cloud complex \citep[cf.][]{2011ApJS..196...18B,2020A&A...644A..64D}. We find that high-intensity ratios of $R_\mathrm{12CO32/13CO21}$ $\sim$10 are distributed toward the shell of W49B in the 10 km s$^{-1}$ cloud. The southwestern edge of the shell (hereafter SW-edge) shows the highest value of $R_\mathrm{12CO32/13CO21}$ $\sim$20, where the radio continuum shell is strongly deformed. On the other hand, the 40 km s$^{-1}$ cloud shows no significant enhancement of $R_\mathrm{12CO32/13CO21}$ toward the SNR shell. The southeast shell in the 60 km s$^{-1}$ cloud also shows higher values of $R_\mathrm{12CO32/13CO21}$, while the regions with high-intensity ratios continuously extend beyond the radio-shell boundary and may not be related to W49B.

\setcounter{figure}{4}
\begin{figure*}[]
\begin{center}
\includegraphics[width=\linewidth,clip]{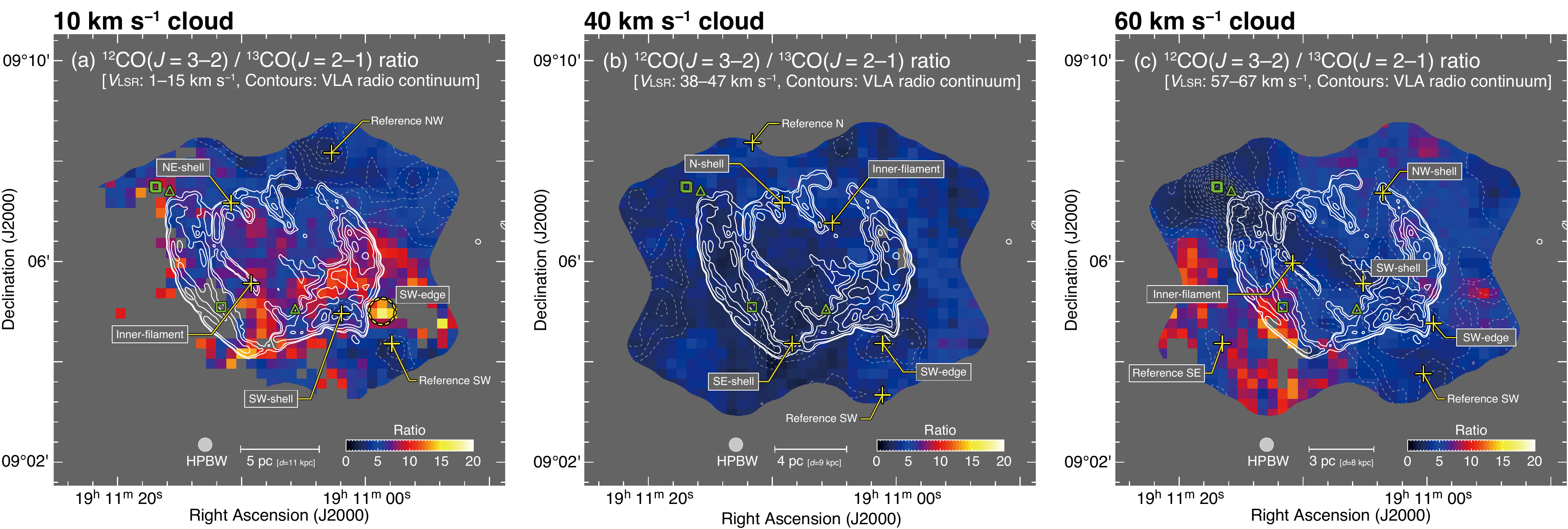}
\caption{Intensity ratio maps of $^{12}$CO($J$ = 3--2) / $^{13}$CO($J$ = 2--1) for (a) the 10~km~s$^{-1}$ cloud, (b) the 40~km~s$^{-1}$ cloud, and (c) the 60~km~s$^{-1}$ cloud. Each data was smoothed to match the beam size 16\farcs6. The velocity range of each cloud and superposed solid contours are the same as shown in Figure \ref{fig1}. Superposed dashed contours indicate $^{13}$CO($J$ = 2--1) integrated intensities, whose lowest contour levels are 3 K km s$^{-1}$ and contour intervals are 1.0 K km s$^{-1}$ for the 10~km~s$^{-1}$ cloud; 0.5 K km s$^{-1}$ for the 40~km~s$^{-1}$ cloud; and 1.5 K km s$^{-1}$ for the 60~km~s$^{-1}$ cloud. The gray areas represent that the $^{12}$CO($J$ = 3--2) and/or $^{13}$CO($J$ = 2--1) data show the lower significance than $\sim$7$\sigma$. Yellow crosses (Reference SW/NW, SW/NE-shell, central-filament) and a dashed circle (SW-edge) discussed in Section \ref{subsec:lvg} are indicated. Green squares and triangles also indicate the positions of \ion{H}{2} regions and IRAS point sources, respectively \citep{1988iras....1.....B,2009A&A...501..539U}.} 
\label{fig5}
\end{center}
\end{figure*}%

\subsection{Physical conditions of the molecular clouds}\label{subsec:lvg}
\begin{figure*}[]
\begin{center}
\includegraphics[width=\linewidth,clip]{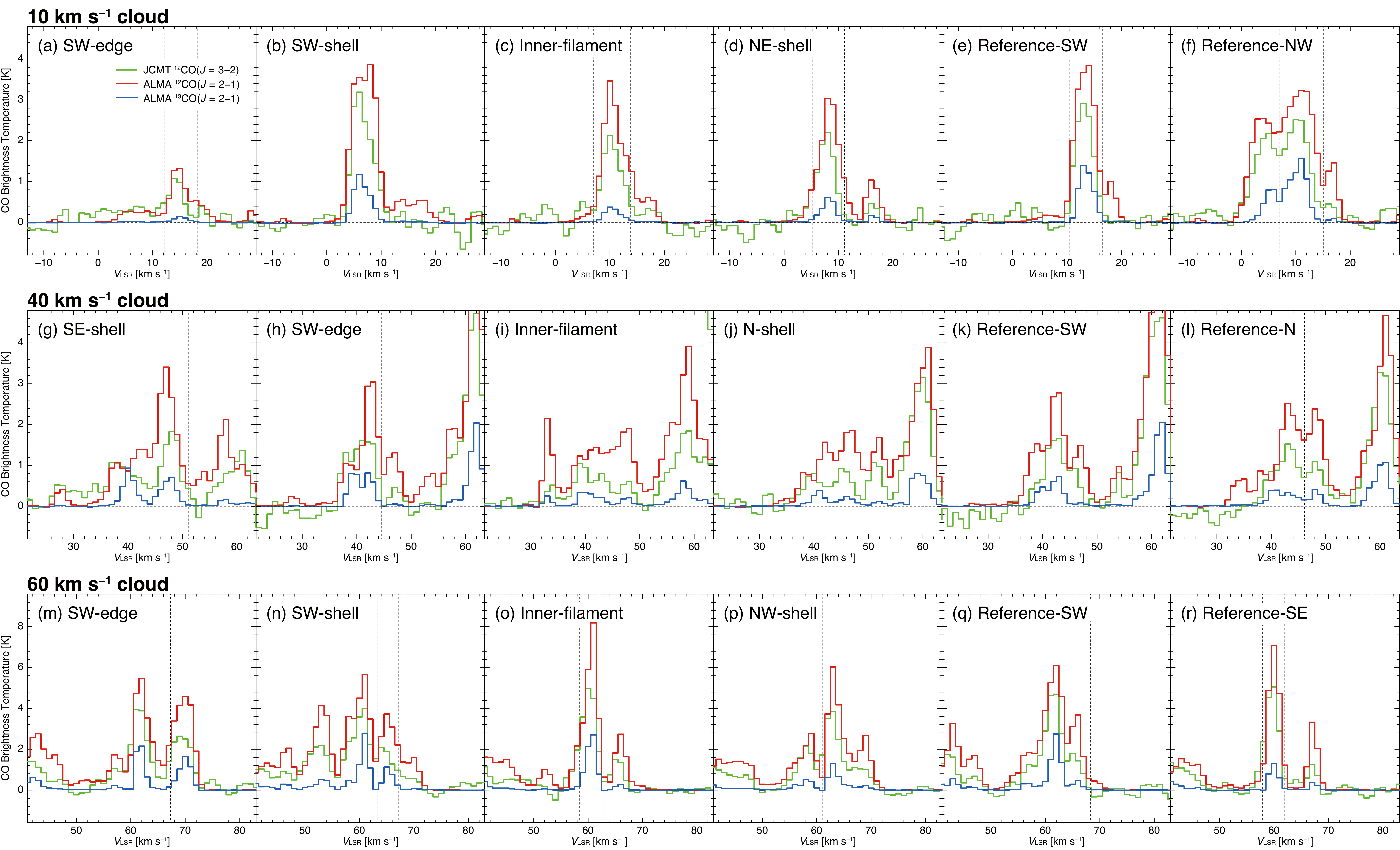}
\caption{CO intensity profiles toward individual CO peaks in (a--f) the 10~km~s$^{-1}$ cloud, (g--l) the 40~km~s$^{-1}$ cloud, and (m--r) the 60~km~s$^{-1}$ cloud. The CO spectra enclosed by vertical lines represent individual CO peaks that are focused on the present study. Each CO data was smoothed to match the beam size 16\farcs6 and the velocity resolution of 1 km s$^{-1}$. In SW-edge spectra of the 10~km~s$^{-1}$ cloud, we combined four pixels around the highest intensity ratio of $^{12}$CO($J$ = 3--2) / $^{13}$CO($J$ = 2--1) as shown in Figure \ref{fig5}a.}
\label{fig6}
\end{center}
\end{figure*}

To reveal physical conditions for each cloud in detail, we performed the Large Velocity Gradient (LVG) analysis \citep[e.g.,][]{1974ApJ...189..441G,1974ApJ...187L..67S}. The LVG analysis can calculate the radiative transfer of molecular line emission, assuming a spherically isotropic cloud with uniform photon escape probability, temperature, and radial velocity gradient of $dv/dr$. Here, $dv$ is the half-width half-maximum of CO line profiles and $dr$ is a cloud radius. We selected six individual CO peaks for each cloud, which are significantly detected both the $^{12}$CO and $^{13}$CO emission lines. Four of them appear to be along the radio continuum shell or filaments (hereafter refer to as ``shell clouds''), and the others were selected as reference which are located outside of the shell (hereafter refer to as ``reference clouds''). CO spectra toward each position are shown in Figure \ref{fig6}. We adopt $dv/dr = {2.5}$ km s$^{-1}$ pc$^{-1}$ for the 10 km s$^{-1}$ SW-shell and the 40 km s$^{-1}$ SE-shell; {1.5} km s$^{-1}$ pc$^{-1}$ for the 60 km s$^{-1}$ SW-{edge}, NW-shell, and Reference-{SW}; and {1.0} km s$^{-1}$ pc$^{-1}$ for the others. We also utilized the abundance ratio of [$^{12}$CO/H$_2$] = $5 \times 10^{-5}$ \citep{1987ApJ...315..621B} and the isotope abundance ratio of [$^{12}$CO/$^{13}$CO] = 49 \citep{1990ApJ...357..477L}.

\begin{figure*}[]
\begin{center}
\includegraphics[width=\linewidth,clip]{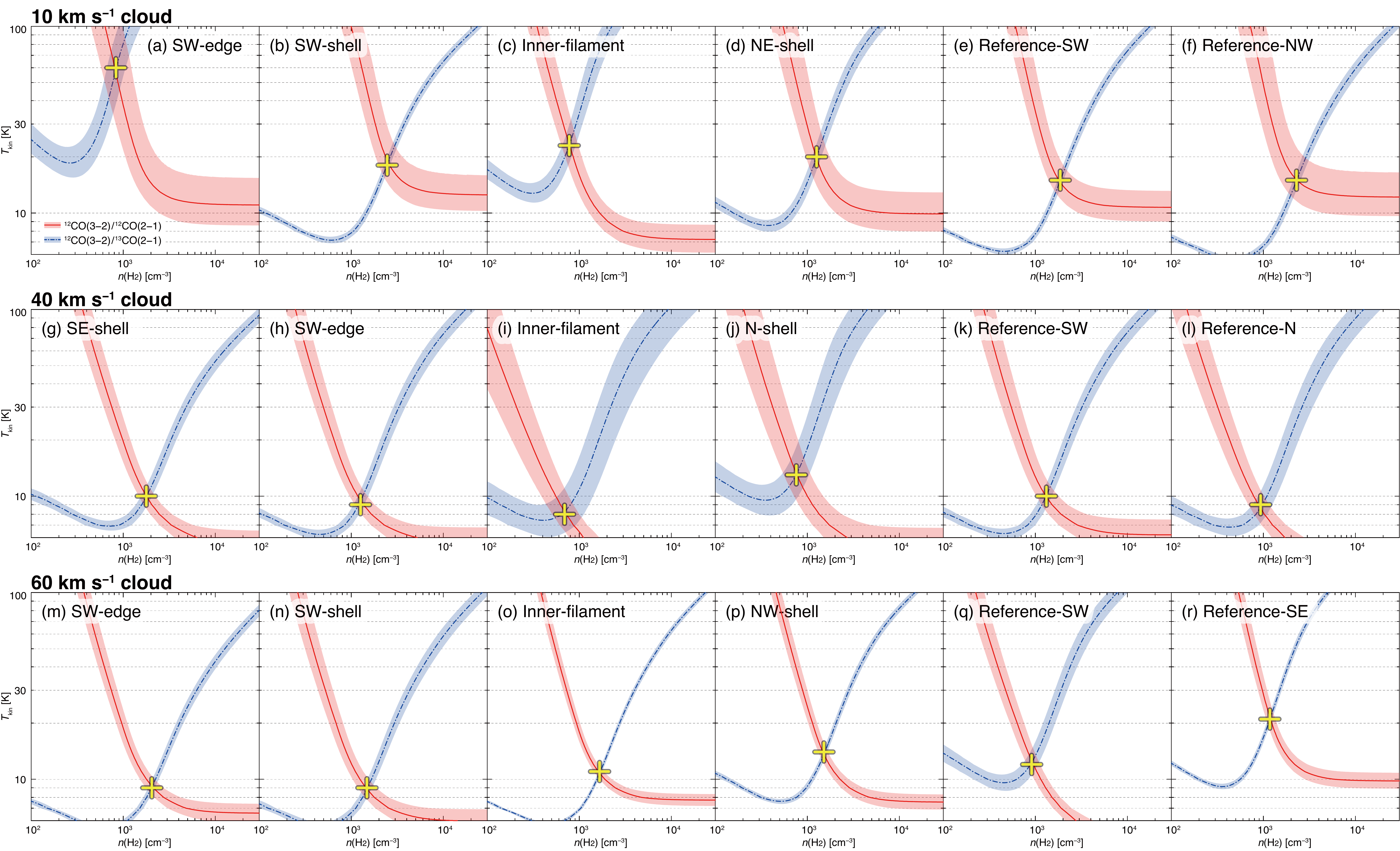}
\caption{LVG results on the number density of molecular hydrogen, $n$(H$_2$), and the kinetic temperature, $T_{\mathrm{kin}}$ for each cloud as shown in Figure \ref{fig6}. The red lines and blue dash-dotted lines indicate the intensity ratios of $^{12}$CO($J$ = 3--2) / $^{12}$CO($J$ = 2--1) and $^{12}$CO($J$ = 3--2) / $^{13}$CO($J$ = 2--1), respectively. The shaded areas in red and blue represent $1\sigma$ error ranges of each intensity ratio. Yellow crosses indicate the best-fit values of $n$(H$_2$) and $T_{\mathrm{kin}}$ for each cloud. The results are summarized in Table \ref{tab1}.}
\label{fig7}
\end{center}
\end{figure*}%

Figure \ref{fig7} shows the LVG results on the number density of molecular hydrogen, $n$(H$_2$), and the kinematic temperature, $T_\mathrm{kin}$, toward the six positions for each cloud. The best-fit values of $n$(H$_2$) and $T_\mathrm{kin}$ are summarized in Table \ref{tab1}. In the 10 km s$^{-1}$ cloud, we find that the shell clouds show $T_\mathrm{kin}$ $\sim$20--{60} K, which are significantly higher than that of the reference clouds ($T_\mathrm{kin} = {15}$ K). By contrast, all shell clouds in both the 40 km s$^{-1}$ and 60 km s$^{-1}$ components show $T_\mathrm{kin}$ $\sim$10 K which are roughly consistent with their reference clouds except for Reference-SE in the 60 km s$^{-1}$ ($T_\mathrm{kin} = {21}$ K). We also note that there is no relation between the number density of molecular hydrogen and kinetic temperature for each cloud.

\section{DISCUSSION}\label{sec:discussion}
\subsection{Molecular Clouds Associated with W49B}\label{subsec:mc}
Previous studies proposed three candidates of molecular clouds which are interacting with the SNR W49B: namely the 10 km s$^{-1}$ cloud, 40 km s$^{-1}$ cloud, and the 60 km s$^{-1}$ cloud \citep{2014ApJ...793...95Z,2016ApJ...816....1K,2020AJ....160..263L}. Their claim is mainly based on three elements: (1) a line-broadening feature of CO emission at $\sim$10 km s$^{-1}$, (2) a wind-bubble like morphology of CO cloud at $\sim$40 km s$^{-1}$, and (3) a central velocity of shocked H$_2$ line emission at $\sim$64 km s$^{-1}$. In this section, we discuss which cloud is the most likely to be associated with W49B in terms of spatial distributions, presence of the expanding gas motion, and physical conditions of CO clouds.

\subsubsection{Spatial Distributions of CO Clouds}
We first emphasize that the 10 km s$^{-1}$ cloud shows a clear spatial correspondence with the radio continuum shell and filaments (see Figure \ref{fig3} and Section \ref{subsec:comprc}). In particular, a majority of CO clouds at $\sim$10 km s$^{-1}$ are located along the outer boundary of the radio continuum shell: $^{13}$CO clouds in the northern shell at 6.6--9.4 km s$^{-1}$, and arc-like $^{12}$CO clouds in the southwestern shell at 12.2--15.0 km s$^{-1}$. Moreover, $^{13}$CO clumps at 3.8--9.4 km s$^{-1}$ spatially coincide well with radio filaments inside the shell. Such spatial correspondence is naturally expected as a result of shock--cloud interactions. According to magneto-hydrodynamical (MHD) simulations, interactions between supernova shocks and clumpy clouds enhance turbulent magnetic field up to $\sim$1 mG on the surface of shocked clouds, where the synchrotron radio/X-ray radiation becomes brighter \citep{2009ApJ...695..825I,2012ApJ...744...71I,2019MNRAS.487.3199C}. This was further supported by several observations toward the Galactic and Magellanic SNRs \citep[][]{2013ApJ...778...59S,2017ApJ...843...61S,2017JHEAp..15....1S,2019ApJ...873...40S,2020ApJ...904L..24S,2018ApJ...863...55Y,2018ApJ...864..161K}, and hence it should not be surprising that shock--cloud interactions with the magnetic field amplification occurred in W49B as well.

However, we cannot rule out the possibility of shock-interaction with the 40 km s$^{-1}$ and 60 km s$^{-1}$ clouds from the spatial comparative studies alone. In fact, molecular clouds at 38.0--40.0 km s$^{-1}$ and 42.0--44.0 km s$^{-1}$ show good spatial correspondences with the southeastern half and southwestern shell of the SNR, respectively (see Figure \ref{fig3}). The 63.0--65.0 km s$^{-1}$ CO map also shows a good anti-correlation with the radio shell, which is not inconsistent with the picture of magnetic field amplification via the shock--cloud interaction.

\subsubsection{Shock and Wind Induced Expanding Gas Motion}
We argue that the cavity-like structures in the position-velocity diagrams at the 10 km s$^{-1}$ and 40 km s$^{-1}$ clouds provide further supports for the shock interactions (see Figures \ref{fig4}a--\ref{fig4}d). Because such cavity-like structure toward a SNR indicates an expanding gas motion, which is thought to be formed by a combination of shock acceleration and strong gas winds from the progenitor system: stellar winds from a high-mass progenitor or disk winds from a progenitor system of post single-degenerate explosion. In the present study, an expanding velocity $\Delta V$ is derived to be $\sim$6 km s$^{-1}$ for the 10 km s$^{-1}$ cloud; and $\sim$3 km s$^{-1}$ for the 40 km s$^{-1}$ cloud. These values are roughly consistent with other Galactic/Magellanic SNRs \citep[e.g.,][]{1990ApJ...364..178K,1991ApJ...382..204K,2017JHEAp..15....1S,2019ApJ...881...85S,2018ApJ...864..161K}. 

It is noteworthy that the two expanding cavities are independent because their $\Delta V$ values are much smaller than the velocity difference of the 10 km s$^{-1}$ and 40 km s$^{-1}$ clouds. Therefore, either expanding shell is located at the same distance with W49B, and the forward shock has been impacted the wind-cavity wall where the shock--cloud interaction occurred. The other expanding shell is likely not associated with W49B. According to \cite{2020PASJ...72L..11S}, there are many relics of fully evolved SNRs in the Galactic plane that cannot be observed by radio continuum, optical, infrared, and X-rays. Because thermal radiation from SNRs has shorter cooling time (below $\sim$10 kyr) than the lifetime of giant molecular clouds \citep[$\sim$10 Myr, e.g.,][]{1999PASJ...51..745F,2009ApJS..184....1K}. The expanding gas motion of the 10 km s$^{-1}$ or 40 km s$^{-1}$ cloud is therefore likely one of such objects that happen to be located along the line of sight.

\subsubsection{Kinetic Temperature of Molecular Clouds}
According to the LVG analysis in Section \ref{subsec:lvg}, the higher kinetic temperature $T_{\mathrm{kin}} \sim$20--{60} K of the shell clouds are seen in the 10 km s$^{-1}$ cloud, suggesting that shock-heating likely occurred. Because the $T_{\mathrm{kin}}$ values are roughly consistent with the previous studies of shock-heated molecular clouds in the vicinity of middle-aged SNRs \citep[e.g.,][]{1998ApJ...505..286S,2012A&A...542L..19G,2013ApJ...768..179Y,2014A&A...569A..81A,2020A&A...644A..64D}. In addition, the presence of high temperature dust components of $45 \pm 4$ K and $151 \pm 20$ K also support the shock-heating scenario \citep{2014ApJ...793...95Z}. Moreover, the bright 24-micron emission is detected in the southwestern shell where SW-edge of the 10 km s$^{-1}$ cloud shows the highest kinetic temperature of $\sim${60} K. It is noteworthy that there are no other extra heating sources such as IRAS point sources or \ion{H}{2} regions toward the shell clouds (see also Figure \ref{fig5}). 

By contrast, all shell clouds in the 40 km s$^{-1}$ and 60 km s$^{-1}$ components show $T_{\mathrm{kin}} \sim$10 K, implying quiescent molecular clouds without any extra-heating processes such as shock heating and stellar radiation. Interestingly, Reference-SE in the 60 km s$^{-1}$ cloud shows warmer temperature of $\sim$20 K, despite the reference cloud is far from the SNR shell. A possible scenario is that a part of the 60 km s$^{-1}$ cloud is located at the tangent point of the Galaxy, and hence the velocity crowding would accumulate diffuse gas and increase the ambient gas temperature \cite[e.g.,][]{2019ApJS..240...14L}. In any case, there is no shock-heated gas in both the 40 km s$^{-1}$ and 60 km s$^{-1}$ clouds.

\begin{deluxetable*}{lccccccc}
\tablecaption{Results of LVG analysis at the 10 km s$^{-1}$, 40 km s$^{-1}$, and 60 km s$^{-1}$ clouds}
\tablehead{
&  \multicolumn{2}{c}{$\mathrm{^{12}CO}$} & &$\mathrm{^{13}CO}$ & & \\
\cline{2-3}\cline{5-5}
 \multicolumn{1}{c}{Name} & $J$~=~3--2 & $J$~=~2--1 &&$J$~=~2--1 & $n$(H${_2}$) & \phantom{0}$T_{\mathrm{kin}}$\phantom{0} \\
 & (K) & (K) &&(K) & \phantom{0}($\times10^3$ cm$^{-3}$)\phantom{0} & (K) \\
 \multicolumn{1}{c}{(1)} & (2) & (3) && (4) & (5) & (6)}
\startdata
\uline{10 km s$^{-1}$ cloud}\phantom{0}\phantom{0}\phantom{0}&&&&&&\\
SW-edge & 1.02 & 1.35 &\phantom{0}& 0.14 & {$0.83^{+0.10}_{-0.05}$} & {$60^{+47}_{-26}$} \\
SW-shell & 3.12 & 3.97 && 1.12 & {$2.45^{+0.50}_{-0.16}$} & ${18}^{{+7}}_{-3\phantom{0}}$ \\
Inner-filament & 2.06 & 3.25 && 0.37 & {$0.78^{+0.03}_{-0.02}$} & {$23^{+9}_{-5\phantom{0}}$} \\
NE-shell & 2.16 & 2.96 && 0.60 & {$1.26^{+0.15}_{-0.09}$} & ${20}^{{+8}}_{-4\phantom{0}}$ \\
Reference-SW & 3.02 & 4.03 && 1.43 & {$1.86^{+0.14}_{-0.08}$} & ${15}^{{+3}}_{-2\phantom{0}}$ \\
Reference-NW & 2.49 & 3.19 && 1.41 & {$2.29^{+0.46}_{-0.20}$} & ${15}^{+6}_{{-3}\phantom{0}}$ \\
\hline
\uline{40 km s$^{-1}$ cloud}\phantom{0}\phantom{0}&&&&&&\\
SE-shell & 1.81 & 3.32 && 0.68 & {$1.78^{+0.08}_{-0.00}$} & $10^{{+2}}_{{-1}\phantom{0}}$ \\
SW-edge & 1.78 & 3.13 && 2.40 & {$1.26^{+0.03}_{-0.03}$} & $\phantom{0}9^{+2}_{-1\phantom{0}}$ \\
Inner-filament & 0.64 & 1.91 && 0.22 & {$0.69^{+0.02}_{-0.00}$} & $\phantom{0}8^{+3}_{-2\phantom{0}}$ \\
N-shell & 0.93 & 5.22 && 0.23 & {$0.76^{+0.05}_{-0.02}$} & $13^{{+6}}_{{-3}\phantom{0}}$ \\
Reference-SW & 1.65 & 2.83 && 0.77 & {$1.32^{+0.09}_{-0.09}$} & $10^{+2}_{-2\phantom{0}}$ \\
Reference-N & 1.05 & 2.40 && 0.42 & $0.93^{+0.03}_{-0.00}$ & $\phantom{0}9^{+2}_{-2\phantom{0}}$ \\
\hline
\uline{60 km s$^{-1}$ cloud}\phantom{0}\phantom{0}&&&&&&\\
SW-edge & 2.78 & 4.61 && 1.62 & $2.04^{+0.05}_{-0.04}$ & $\phantom{0}9^{+1}_{-1\phantom{0}}$ \\
SW-shell & 2.09 & 3.66 && 1.19 & $1.48^{+0.03}_{-0.03}$ & $\phantom{0}9^{+1}_{-1\phantom{0}}$ \\
Inner-filament & 5.22 & 7.93 && 2.82 & {$1.66^{+0.08}_{-0.04}$} & $11^{{+1}}_{-1\phantom{0}}$ \\
NW-shell & 3.98 & 6.13 && 1.31 & {$1.51^{+0.04}_{-0.03}$} & ${14}^{+1}_{{-2}\phantom{0}}$ \\
Reference-SW & 1.76 & 3.64 && 0.42 & {$0.91^{+0.02}_{-0.00}$} & $12^{{+3}}_{{-1}\phantom{0}}$ \\
Reference-SE & 5.20 & 7.15 && 1.33 & $1.17^{{+0.06}}_{{-0.02}}$ & {$21^{+3}_{-2\phantom{0}}$} \\
\enddata
\label{tab1}
\tablecomments{Col. (1): Region name for each cloud. Cols. (2)--(3): Radiation temperature for each line emission derived by the least-squares fitting using a single Gaussian function. Col. (4): Number density of molecular hydrogen. Col. (5): Kinetic temperature.}
\vspace*{-0.5cm}
\end{deluxetable*}

\subsubsection{Final Decision and Consistency with Previous Studies}\label{sec:final}
In conclusion, we claim that the 10 km s$^{-1}$ cloud is the one most likely associated with W49B in terms of its spatial distribution, kinetics, and physical conditions. This velocity is consistent not only with the line-broadening measurements by \cite{2016ApJ...816....1K}, but also with the latest \ion{H}{1} absorption measurement toward W49B by \cite{2018AJ....155..204R}. In this case, the kinematic distance of W49B is slightly revised to $11.0 \pm 0.4$ kpc assuming the Galactic rotation curve model of \cite{1993A&A...275...67B} and the latest Galactic parameters of $R_0 = 7.92$ kpc and $\Theta_0 = 227$ km s$^{-1}$ \citep{2020PASJ...72...50V}. This value is also roughly consistent with the previous distance to W49B of $11.3 \pm 0.4$ kpc derived by \ion{H}{1} absorption \citep{2018AJ....155..204R}.

On the other hand, there is large gap in radial velocities between the 10 km s$^{-1}$ cloud and shocked H$_2$ line emission at $\sim$64 km s$^{-1}$ \citep{2020AJ....160..263L}. We argue that this inconsistency should not be a problem considering the excitation condition for each line emission. In general, the CO line emission at 2.6 mm (also known as $^{12}$CO $J$ = 1--0 transition line) can trace a bulk mass of molecular cloud with low kinetic temperature of $\sim$10 K. On the other hand, the supernova-shocked H$_2$ line emission traces only a small portion of molecular cloud which is highly excited into $\sim$2000--3000 K \citep[e.g.,][]{1994ApJ...427..777M,2020AJ....160..263L}. In W49B, the CO-traced molecular cloud mass is $\sim$$2.7 \times 10^4$ $M_{\sun}$ for the 10 km s$^{-1}$ cloud, whereas the mass of shocked H$_2$ is only 14--550 $M_{\sun}$ \citep{2007ApJ...654..938K}. This indicates that the shocked H$_2$ mass is only $\sim$2\% of the CO-traced molecular cloud mass at most. Considering the momentum conservation between the SNR shocks and interacting clouds, the shocked H$_2$ component is more easily accelerated than the CO-traced molecular cloud. We therefore argue that the velocity inconsistency between the CO cloud and shocked H$_2$ component is caused by the difference of their excitation conditions and masses.

In any case, the physical interaction of the 10 km s$^{-1}$ cloud with W49B means that the bulk mass of molecular clouds is concentrated in the northwestern half of W49B, not in the southwest shell. This is consistent with the hydrogen density maps derived by \cite{2018AA...615A.150Z}, who found higher plasma densities in the west of W49B by the X-ray spectral modeling. The inhomogeneous gas distribution will significantly affect to understand the origins of recombining plasma and gamma-rays from W49B. We will discuss later them the latter Sections \ref{subsec:rp} and \ref{subsec:wp}.

\subsection{A Detailed Comparison with X-Rays}\label{subsec:xcomp}
\begin{figure}[]
\begin{center}
\includegraphics[width=\linewidth,clip]{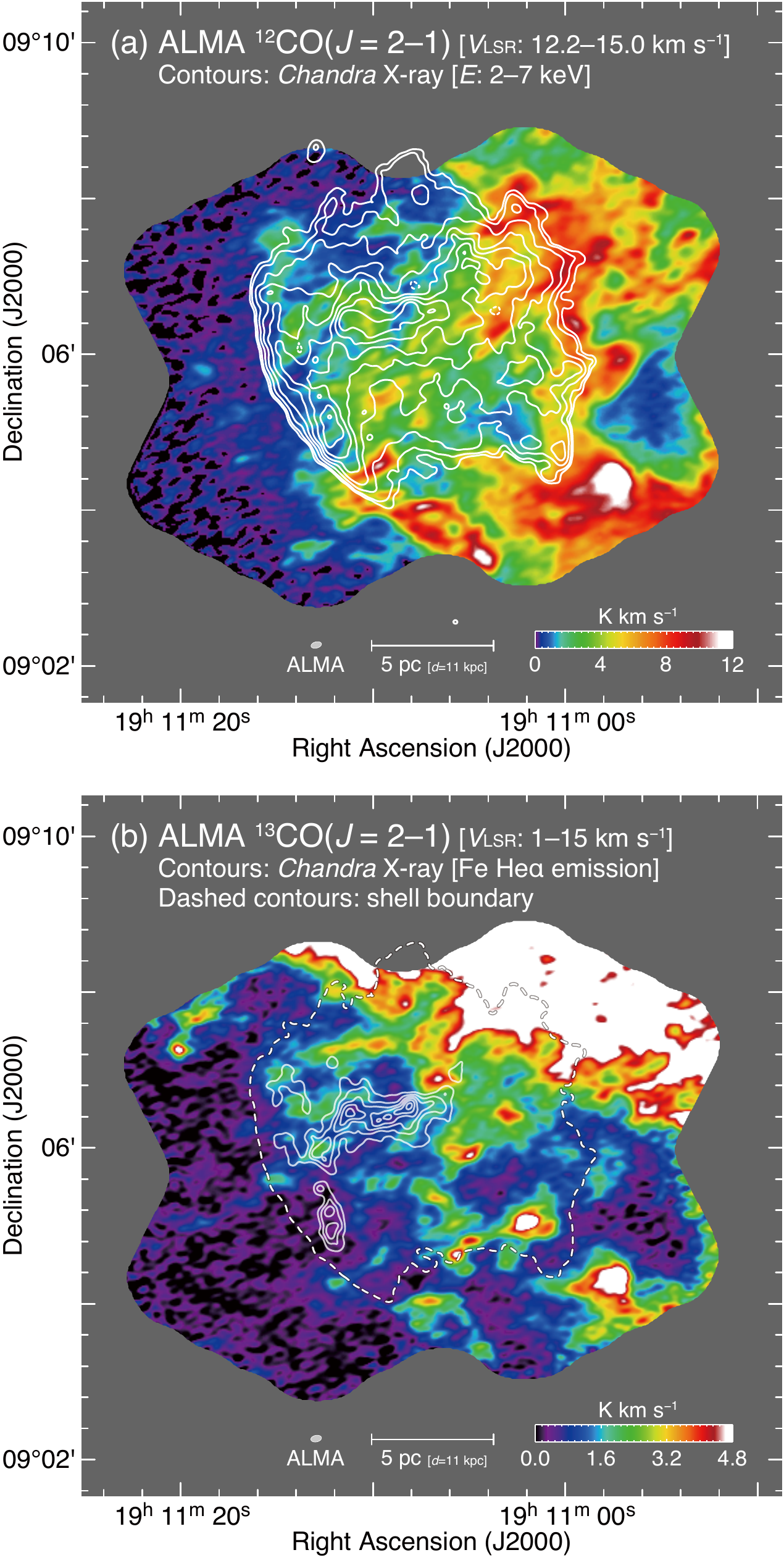}
\caption{(a) Integrated intensity maps of ALMA ACA $^{12}$CO($J$ = 2--1) superposed on the {\it{Chandra}} X-ray intensity contours in the energy band of 2--7 keV. The integration velocity range of CO is from 12.2 to 15.0 km s$^{-1}$. The contour levels are 0.3, 0.4, 0.7, 1.2, 1.9, 2.8, 3.9, and $5.2 \times 10^{-7}$ photon cm$^{-2}$ s$^{-1}$. (b) Integrated intensity maps of ALMA ACA $^{13}$CO($J$ = 2--1) superposed on the continuum-subtracted Fe He$\alpha$ emission. The integration velocity range of CO is from 1.0 to 15.0 km s$^{-1}$. The contour levels are 0.5, 0.8, 1.1, 1.4, 1.7, and $2.0 \times 10^{-7}$ photon cm$^{-2}$ s$^{-1}$.}
\label{fig8}
\end{center}
\end{figure}

To reveal a physical relation between the 10 km s$^{-1}$ cloud and X-ray radiation, we here compare the CO distributions with {\it{Chandra}} X-ray images\footnote{Because the shell boundary of X-rays is almost similar to that of radio continuum except for the western half, we here only present a spatial comparison with the CO map of 12.2--15.0 km s$^{-1}$ which is bright in the western part of the shell.}. Figure \ref{fig8}a shows the ALMA ACA $^{12}$CO($J$ = 2--1) integrated intensity overlaid with the {\it{Chandra}} X-ray contours at the energy band of 2--7 keV. In the integration velocity range of 12.2--15.0 km s$^{-1}$, CO clouds are perfectly along with the zigzag pattern of the western X-ray shell, indicating that the shock ionization occurred. Note that the spatial correspondence is also seen in the X-ray image at 4.2--5.5 keV, which are mostly free from the interstellar absorption. We thus suggest that the shockwave was strongly decelerated and deformed in the western shell along the dense clouds, whereas the eastern shell was freely expanded with a smooth shape of the forward shock. This also indicates that the shock velocity of the eastern shell is faster than that of the western shell. Further proper motion studies might be able to reveal the velocity difference in the east--west direction.

Figure \ref{fig8}b shows an overlay map of the $^{13}$CO($J$ = 2--1) intensity image and the continuum-subtracted Fe He$\alpha$ emission in white contours. To compare the Fe-rich ejecta with the total amount of dense clouds, we use $^{13}$CO with the whole velocity range of 1.0--15.0 km s$^{-1}$. Although the elongated structure of Fe-rich ejecta is believed to be related to a bipolar/jet-driven Type Ib/Ic explosion \citep[][]{2007ApJ...654..938K,2013ApJ...777..145L}, the Fe-rich ejecta is mainly located on the void of dense molecular clouds. Moreover, almost Fe-rich ejecta is surrounded by dense molecular clumps. We argue that this situation is consistent with the supernova explosion inside a barrel-shaped cavity which was proposed by \cite{2018AA...615A.150Z}. The authors revealed that an enhancement of cool plasma component along the Fe-rich ejecta (or the void of dense clouds) was observed by a spatially resolved X-ray spectroscopy (see Figure 4 in \citeauthor{2018AA...615A.150Z}~\citeyear{2018AA...615A.150Z}). Following the proposed scenario, the forward shock was freely expanded in the low-density medium at the beginning, and then suddenly encountered with the dense gaseous materials traced by $^{13}$CO line emission and/or cool plasma component. Since the shock--cloud interaction generates multiple reflected (or inward) shocks, the Fe-rich ejecta is efficiently heated up at higher densities toward the center of the SNR \citep[see also][]{2019ApJ...873...40S}. The X-Ray Imaging and Spectroscopy Mission \citep[XRISM,][]{2020arXiv200304962X} will provide us with further understanding of shock-interactions through a detailed spatial comparison between the X-ray derived ionic properties and CO clouds.

\begin{figure*}[]
\begin{center}
\includegraphics[width=\linewidth,clip]{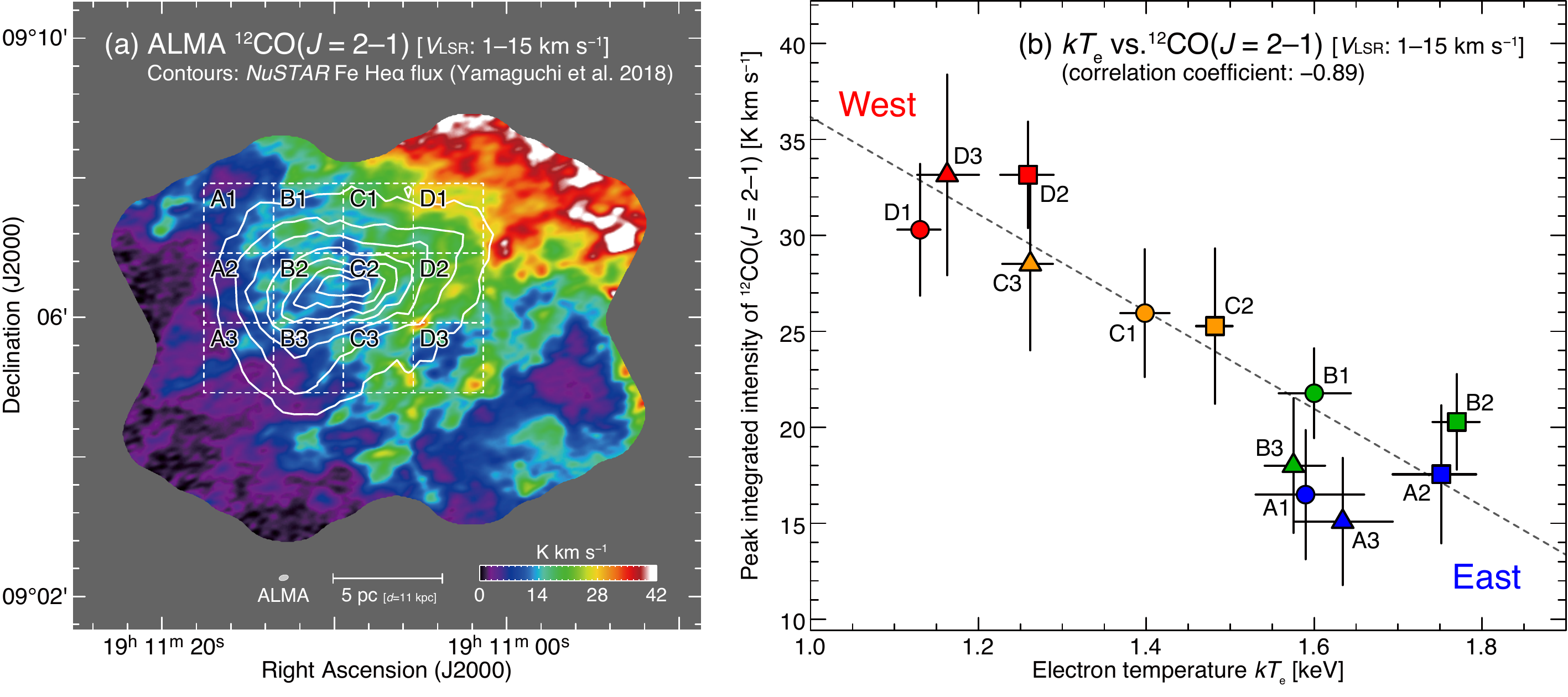}
\caption{(a) Integrated intensity maps of ALMA ACA $^{12}$CO($J$ = 2--1) for the 10~km~s$^{-1}$ cloud superposed on {\it{NuSTAR}} Fe He$\alpha$ flux contours \citep{2018ApJ...868L..35Y}. The integration velocity range and contour levels are the same as shown in Figures \ref{fig2}a and \ref{fig6}, respectively. The bashed 12 boxes indicate $1' \times 1'$ regions used for the spatially resolved spectral analysis in \cite{2018ApJ...868L..35Y} and deriving CO averaged integrated intensities in Figure \ref{fig9}b. (b) Scatter plot between the electron temperature $kT_\mathrm{e}$ \citep{2018ApJ...868L..35Y} and peak integrated intensities of $^{12}$CO($J$ = 2--1) for box regions A1--3, B1--3, C1--3, and D1--3 as shown in Figure \ref{fig9}a. Error bars of CO and $kT_\mathrm{e}$ represent standard division of CO integrated intensity and 1$\sigma$ confidence level for each box region. The dashed line indicates the linear regression applying the least squares method.}
\label{fig9}
\end{center}
\end{figure*}%

\subsection{Origin of the High-Temperature Recombining Plasma in W49B}\label{subsec:rp}
It is a long-standing question how the recombining (overionized) plasma is formed in SNRs since its discovery in 2002 \citep[IC443,][]{2002ApJ...572..897K}. Subsequent detailed X-ray spectroscopic observations revealed that nearly 20 SNRs show the overionized state (e.g., W49B, \citeauthor{2009ApJ...706L..71O}~\citeyear{2009ApJ...706L..71O}; G359.1$-$0.5, \citeauthor{2011PASJ...63..527O}~\citeyear{2011PASJ...63..527O}; W28, \citeauthor{2012PASJ...64...81S}~\citeyear{2012PASJ...64...81S}; W44, \citeauthor{2012PASJ...64..141U}~\citeyear{2012PASJ...64..141U}; G346.6$-$0.2, \citeauthor{2013PASJ...65....6Y}~\citeyear{2013PASJ...65....6Y}; 3C 391, \citeauthor{2014ApJ...790...65E}~\citeyear{2014ApJ...790...65E}; CTB 37A, \citeauthor{2014PASJ...66....2Y}~\citeyear{2014PASJ...66....2Y}; G290.1$-$0.8, \citeauthor{2015PASJ...67...16K}~\citeyear{2015PASJ...67...16K}; LMC N49, \citeauthor{2015ApJ...808...77U}~\citeyear{2015ApJ...808...77U}; Kes 17, \citeauthor{2016PASJ...68S...4W}~\citeyear{2016PASJ...68S...4W}; G166.0$+$4.3, \citeauthor{2017PASJ...69...30M}~\citeyear{2017PASJ...69...30M}; 3C400.2, \citeauthor{2017ApJ...842...22E}~\citeyear{2017ApJ...842...22E}; LMC N132D, \citeauthor{2018ApJ...854...71B}~\citeyear{2018ApJ...854...71B}; HB21, \citeauthor{2018PASJ...70...75S}~\citeyear{2018PASJ...70...75S}; CTB1, \citeauthor{2018PASJ...70..110K}~\citeyear{2018PASJ...70..110K}; Sagittarius A East, \citeauthor{2019PASJ...71...52O}~\citeyear{2019PASJ...71...52O}; G189.6+3.3, \citeauthor{2020PASJ...72...81Y}~\citeyear{2020PASJ...72...81Y}, see also a review by \citeauthor{2020AN....341..150Y}~\citeyear{2020AN....341..150Y}). However, the physical origin of recombining plasmas is still under debate.
 
Since the recombining plasma is characterized by higher ionization temperature $kT_\mathrm{i}$ than the electron temperature $kT_\mathrm{e}$, rapid electron cooling or increasing ionization state is needed to produce the plasma state. Three scenarios have been proposed to explain the origin of recombining plasmas in SNRs, called adiabatic cooling, thermal conduction, and photoionization scenarios. In the adiabatic cooling (a.k.a. rarefaction) scenario, rapid electron cooling occurs when the shockwaves breakout from a dense ISM (e.g., CSM) into a much less dense medium \citep[e.g.,][]{1989MNRAS.236..885I,1994ApJ...437..770M,2018ApJ...868L..35Y}. In the thermal conduction scenario, such rapid electron cooling is caused by interactions between the shockwaves and cold dense clouds through thermal conduction \citep[e.g.,][]{2002ApJ...572..897K,2017PASJ...69...30M,2017ApJ...851...73M,2018PASJ...70...35O,2020ApJ...890...62O}. On the other hand, the photoionization scenario proposes that an external X-ray radiation or low-energy CRs) increase the ionization state via photoionization \citep[e.g.,][]{2013ApJ...773...20N,2019PASJ...71...52O,2019PASJ...71...37H}. Because photoionization can be seen in limited environments such as near the Galactic center or a SNR with strong \ion{Fe}{1} K$\alpha$ emission, the adiabatic cooling and thermal conduction scenarios are thought to be the formation mechanisms of recombining plasmas in most SNRs.

The origin of recombining plasma in W49B has been discussed in the past decade. The thermal conduction scenario was initially proposed by \cite{2005ApJ...631..935K}, whereas the adiabatic cooling scenario is more favored in the subsequent studies \citep{2010A&A...514L...2M,2013ApJ...777..145L,2011MNRAS.415..244Z,2018ApJ...868L..35Y}. Because the recombining plasma in W49B shows a positive correlation between the ionization timescale $n_\mathrm{e}t$ and $kT_\mathrm{e}$. Further, there is no correlation between the
plasma condition and ambient clouds traced by near infrared emission \citep{2018ApJ...868L..35Y}. This trend is in contrast to what is observed in W44 (see also \citeauthor{2020ApJ...890...62O}~\citeyear{2020ApJ...890...62O} and a review in \citeauthor{2020AN....341..150Y}~\citeyear{2020AN....341..150Y}). On the other hand, most recent X-ray studies presented that the X-ray spectra from W49B are reproduced by two ejecta components (low- and high-temperature plasma). The authors proposed thermal conduction scenarios especially for the high-temperature recombining plasma in W49B, considering the conduction timescale \citep{2020ApJ...893...90S,2020ApJ...903..108H}. Note that \cite{2020ApJ...903..108H} argued that thermal conduction is a possible origin of recombining plasma in the eastern regions of W49B because dense molecular clouds are thought to be associated in the southwestern shell \citep{2007ApJ...654..938K,2014ApJ...793...95Z}. In this section, we argue that the origin of the high-temperature recombining plasma in W49B can be understood as the thermal conduction scenario considering the CO-traced interacting molecular clouds in W49B.

\begin{deluxetable*}{lcccccl}[]
\tablecaption{Comparison of Physical Properties in Eleven Gamma-Ray SNRs}
\tablehead{
\multicolumn{1}{c}{Name}  & Distance & Diameter & Age & $n_\mathrm{p}$ & $W_\mathrm{p}$ & \multicolumn{1}{c}{References} \\
 & (kpc) & (pc) & (kyr) & (cm$^{-3}$) & ($10^{49}$~erg) & \\
 \multicolumn{1}{c}{(1)} & (2) & (3) & (4) & (5) & (6) & \multicolumn{1}{c}{(7)}}
\startdata
RX~J1713.7$-3946$ & 1.0 & 18 & 1.6 & 130 & \phantom{00}$0.16^{+0.07}_{-0.08}$ & \cite{2012ApJ...746...82F} \\
RX~J0852.0$-$4622 & \phantom{0z}0.75$^\mathrm{a}$ & 24 & \phantom{z}1.7$^\mathrm{a}$ & 100 &  \phantom{00}$0.07^{+0.02}_{-0.02}$  & \cite{2017ApJ...850...71F} \\
RCW~86 & 2.5 & 30 & 1.8 & \phantom{0}75 & \phantom{00}$0.11^{+0.01}_{-0.01}$ & \cite{2019ApJ...876...37S} \\
HESS~J1731$-$347 & 5.7 & 44 & 4.0&  \phantom{0}60 & \phantom{00}$0.66^{+0.22}_{-0.22}$ & \cite{2014ApJ...788...94F} \\
G39.2$-$0.3 & 6.2 & 14 & \phantom{0}\phantom{0zw}$5.0^{+2.0}_{-2.0}$$^\mathrm{b}$ & 400 & $3.2^{+1.1}_{-0.8}$ & \cite{2020MNRAS.497.3581D} \\
W49B & 11.0\phantom{0} & 16 &  \phantom{0}\phantom{0zw}$6.0^{+1.0}_{-1.0}$$^\mathrm{c}$ & 650 & $2.1^{+1.1}_{-0.6}$  & This work \\
Kes~79 & 5.5 & 16 & \phantom{0zw}$8.3^{+0.5}_{-0.5}$ & 360 & 0.5\phantom{0zw} & \cite{2018ApJ...864..161K} \\
W44 & \phantom{0}3.0$^\mathrm{d}$ & 27 & 20.0$^\mathrm{e}$& 200 & 1.0\phantom{0zw} &  \cite{2013ApJ...768..179Y} \\
IC443 & \phantom{i}1.5$^\mathrm{f}$ & 20 & \phantom{0iiiii}$25.0^{+5.0}_{-5.0}$$^\mathrm{g}$  & 680 & 0.09\phantom{ccz} & \cite{2021MNRASsubmitted} \\
\hline
LMC N132D & 50.0\phantom{0} & 25 & \phantom{00zw}$2.5^{+0.2}_{-0.2}$$^\mathrm{h}$ & $< 2000$& $> 0.5$ & \cite{2020ApJ...902...53S} \\
LMC N63A & 50.0\phantom{0} &  18 & \phantom{00zii}$3.5^{+1.5}_{-1.5}$$^\mathrm{i}$ & 190 & $0.9^{+0.5}_{-0.6}$ & \cite{2019ApJ...873...40S} \\
\enddata
\label{tab2}
\tablecomments{Col. (1): Name of SNRs. Col. (2): Distance to SNRs in units of kpc. Col. (3): Diameter of SNRs in units of pc. Col. (4): Age of SNRs in units of kyr. Col. (5): Averaged number density of total interstellar protons $n_\mathrm{p}$ in units of cm$^{-3}$. Col. (6): Total energy of cosmic-ray protons $W_\mathrm{p}$ in units of 10$^{49}$ erg. Col. (7): References for CO/\ion{H}{1} derived $n_\mathrm{p}$ and $W_\mathrm{p}$ for each SNR. Other specific references are also shown as follows: $^\mathrm{a}$\cite{2008ApJ...678L..35K}, $^\mathrm{b}$\cite{2011ApJ...727...43S}, $^\mathrm{c}$\cite{2018AA...615A.150Z}, $^\mathrm{d}$\cite{1975AA....45..239C}, $^\mathrm{e}$\cite{1991ApJ...372L..99W}, $^\mathrm{f}$\cite{2003AA...408..545W}, $^\mathrm{g}$\cite{2008AJ....135..796L}; \cite{2001ApJ...554L.205O}, $^\mathrm{h}$\cite{2020ApJ...894...73L}, and $^\mathrm{i}$\cite{1998ApJ...505..732H}.}
\end{deluxetable*}

Figure \ref{fig9}a shows the $^{12}$CO($J$ = 2--1) integrated intensity map of the 10 km s$^{-1}$ cloud superposed on the {\it{NuSTAR}} Fe He$\alpha$ flux contours \citep{2018ApJ...868L..35Y}. The twelve $1' \times 1'$ boxes indicate regions used for spatially resolved spectral analysis using {\it{NuSTAR}} by \cite{2018ApJ...868L..35Y}. The authors fitted X-ray spectra above 3~keV using a single temperature model, because the energy band is dominated by the high-temperature plasma. We note that the CO integrated intensities are significantly changed region to region, which shows an intensity gradient from the southeast to the northwest as mentioned in Section \ref{sec:results}.

Figure \ref{fig9}b shows a scatter plot between $kT_\mathrm{e}$ of the high-temperature recombining plasma \citep{2018ApJ...868L..35Y} and peak integrated intensity of $^{12}$CO($J$ = 2--1) line emission for each box. We find a clear negative correlation between the two. More precisely, $kT_\mathrm{e}$ values in high-temperature plasma are increasing from the west (cloud rich) regions to the east (cloud poor) regions. This is consistent with the thermal conduction scenario: rapid electron cooling occurred in cold/dense cloud rich regions. Note that this finding will not rule out the adiabatic cooling scenario in W49B. In fact, the X-ray spectra from W49B are reproduced two ejecta components: the low-temperature recombining plasma favors the adiabatic cooling scenario whereas the high-temperature component is likely produced by the thermal conduction \citep{2020ApJ...893...90S,2020ApJ...903..108H}. In other words, both the thermal conduction and adiabatic cooling processes coexist in W49B.

Finally, we discuss the reason why our conclusion---the thermal conduction origin of the high-temperature plasma---is different from some previous studies. One of the most important issues is the previous evaluation of the ISM interacting with W49B. Almost all previous studies used the shocked H$_2$ distribution as the bulk mass of the ISM. However, as discussed in Section \ref{sec:final}, the shocked H$_2$ mass is only $\sim$2\% of the CO-traced molecular cloud mass. Because the shocked H$_2$ map is bright in the southeast, most of researchers believed that the southeast shell is interacting with dense molecular clouds and the ISM mass of east is higher than that of west. Some previous studies therefore concluded that the lower plasma temperature in west was caused by the adiabatic cooling process \citep[e.g.,][]{2020ApJ...903..108H,2018ApJ...868L..35Y}. By contrast, it is noteworthy that \cite{2018AA...615A.150Z} suggested that molecular cloud density is higher in the west than the east, which is compatible with our ALMA results. We also note there are different interpretations for the $n_\mathrm{e}t$ variation in the high-temperature plasma. \cite{2018ApJ...868L..35Y} used $n_\mathrm{e}t$ as a proxy for electron density, assuming a uniform time since heating and uniform initial temperature. The former contrasts with the results of \cite{2020ApJ...903..108H} and \cite{2018AA...615A.150Z}, who found higher recombination ages in the east than the west. The positive correlation between $n_\mathrm{e}t$ and $kT_\mathrm{e}$ in W49B may have to be reconsidered. In any case, we emphasize that the proper evaluation of the ISM surrounding an SNR is essential to understand the origin of recombining plasma correctly. Further detailed comparative studies of CO based molecular cloud properties and X-ray spectroscopic results are needed to better understand the origin of recombining plasma in SNRs.

\begin{figure*}
\begin{center}
\includegraphics[width=130mm,clip]{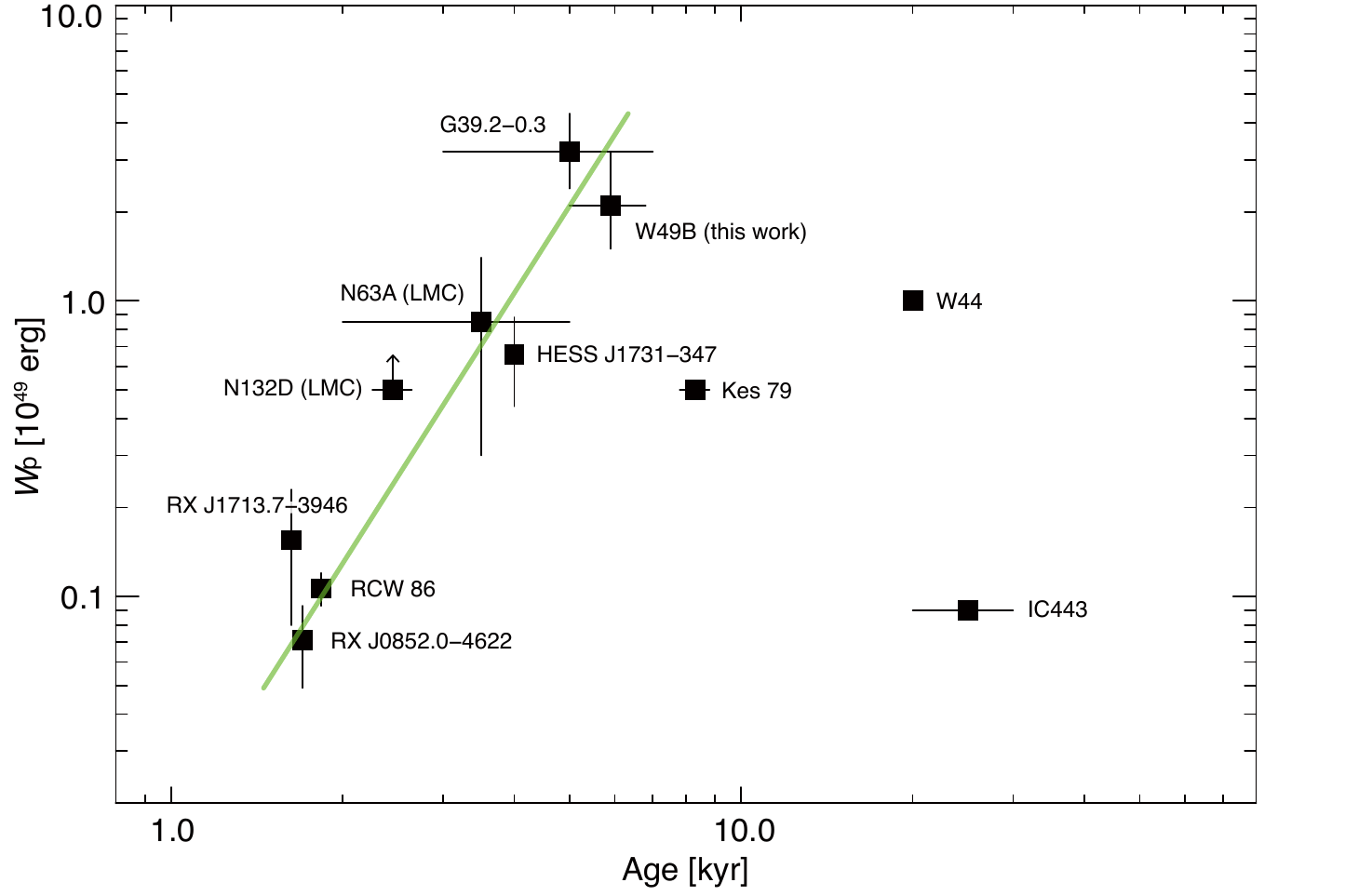}
\caption{Scatter plot between the age of SNRs and the total energy of cosmic-ray protons $W_\mathrm{p}$. The data points and references are summarized in Table \ref{tab2}. The green solid line indicates the linear regression of the double-logarithmic plot applying the least squares method.}
\label{fig10}
\end{center}
\end{figure*}%

\subsection{Total Energy of Cosmic-Ray Protons}\label{subsec:wp}
It is a hundred-year problem how CRs, mainly comprising relativistic protons, are accelerated in interstellar space. SNRs are believed to be acceleration sites for Galactic CRs below $\sim$3~PeV through the diffusive shock acceleration \citep[DSA, e.g.,][]{1978MNRAS.182..147B,1978ApJ...221L..29B}. A conventional value of the total energy of CRs accelerated in an SNR is thought to be $\sim$$10^{49}$--$10^{50}$~erg, corresponding to $\sim$1--10\% of typical kinematic energy released by a supernova \citep[$10^{51}$~erg, e.g.,][]{2013ASSP...34..221G,2019AJ....158..149L}. One of the foremost challenges is to validate these predictions experimentally.

Gamma-ray and radio-line observations hold a key to understand the acceleration of CRs in SNRs. Gamma-rays from SNRs are produced by two different mechanisms: hadronic and leptonic processes \citep[e.g.,][]{1994A&A...285..645A,1994A&A...287..959D}. For the hadronic process, CR proton--interstellar proton interaction creates a neutral pion that quickly decays into two gamma-ray photons (hadronic gamma-ray). For the leptonic scenario, a CR electron energizes a low-energy photon into gamma-ray energy via inverse Compton scattering, in addition to produce gamma-rays through non-thermal Bremsstrahlung (leptonic gamma-ray). To confirm the acceleration of relativistic protons, the main component of CRs, it is crucial to detect the characteristic spectral feature of hadronic gamma-rays with a cut-off at a few GeV known as pion-decay bump \citep[e.g.,][]{2011ApJ...742L..30G,2013Sci...339..807A}. In addition, a good spatial correspondence between gamma-rays and interstellar protons provides an alternative support for the CR proton acceleration \citep[e.g.,][]{2003PASJ...55L..61F,2012ApJ...746...82F,2017ApJ...850...71F,2008A&A...481..401A,2013ApJ...768..179Y,2019ApJ...876...37S}, because the hadronic gamma-ray flux is proportional to the total energy of CR protons and the number density of interstellar protons. Note that the interstellar protons are mainly either neutral molecular and atomic hydrogen traced by CO and \ion{H}{1} radio-lines, respectively. Adopting the number density of targeted interstellar protons, the total energy of CR protons was estimated to be $W_\mathrm{p}\sim$10$^{48}$--$10^{49}$~erg toward a dozen gamma-ray SNRs. However, it is still under debate which parameters are important to understand the variety of observed (or in-situ) $W_\mathrm{p}$ values. To better understand the origins of CR protons and variety of $W_\mathrm{p}$, we need more samples as well as detailed gamma-ray and radio-line studies for SNRs.

W49B is thought to be one of the CR proton accelerators because of the detection of hadron-dominant gamma-rays with a pion-decay bump \citep{2018A&A...612A...5H}. In fact, the best-fit position of GeV gamma-ray detected by {\it{Fermi}} Large Area Telescope (LAT) is the edge of the SE shell, where the bright radio continuum, shocked H$_2$ emission, and the dense molecular clouds are located. To obtain the total energy of CR protons in W49B, we first estimate the number density of interstellar protons interacting with the SNR. Using the equations (1) and (2), the averaged number density of interstellar protons in molecular form is estimated to be $\sim$$650 \pm 200$ cm$^{-3}$ assuming a shell radius of 8 pc and a thickness of 3 pc \citep[e.g.,][]{1994ApJ...437..705M}. The error is derived as the typical uncertainty of the CO-to-H$_2$ conversion factor of $\sim$30\% \cite[cf.][]{2013ARA&A..51..207B}. Additionally, the interstellar protons in atomic form are neglectable in W49B because the derived column density of atomic hydrogen is significantly lower than that of molecular hydrogen \citep[see][]{2001ApJ...550..799B}. The similar situation is also seen in other middle-aged SNRs \citep[e.g., W44,][]{2013ApJ...768..179Y}. We therefore adopt the number density of interstellar protons $n_\mathrm{p}$ to be $\sim$$650 \pm 200$ cm$^{-3}$.

According to \cite{2018A&A...612A...5H}, the total energy of CR protons $W_\mathrm{p}$ is written as
\begin{eqnarray}
W_\mathrm{p} \sim2.0\mathrm{-}2.2 \times 10^{49} (n_\mathrm{p} / 650\; \mathrm{cm^{-3}})^{-1}\; (d / 11\; \mathrm{kpc})^{-2}  \;\; \mathrm{erg},\;\;\;\;\;
\label{eq3}
\end{eqnarray}
where $d$ is the distance to the SNR. Adopting $n_\mathrm{p} = 650$ cm$^{-3}$ and $d = 11$ kpc, we than obtain $W_\mathrm{p} \sim$$2 \times 10^{49}$ erg, corresponding to $\sim$2\% of the typical kinematic energy released by a supernova explosion. Table \ref{tab2} compares physical properties of eleven gamma-ray SNRs including W49B. Here, all values of $n_\mathrm{p}$ and $W_\mathrm{p}$ were derived from CO/\ion{H}{1} radio-line observations. We find that $W_\mathrm{p}$ in W49B is roughly consistent with that in other gamma-ray bright SNRs located in our Galaxy or the Large Magellanic Cloud (LMC). In addition, it is noteworthy that young SNRs RX~J1713.7$-$3946, RX~J0852.0$-$4622 (a.k.a. Vela Jr.), and RCW~86 as well as an evolved SNR IC~443 show the lowest values of $W_\mathrm{p} \sim$$10^{48}$ erg, while the others hold higher values of $W_\mathrm{p} \sim$$10^{49}$ erg.

To better understand the trend, we plot $W_\mathrm{p}$ values as a function of the age of SNRs. Figure \ref{fig10} shows a scatter plot between the age of SNRs and $W_\mathrm{p}$. We find a positive correlation between two parameters in the SNRs with a young age less than $\sim$6000 yrs, suggesting that in-situ values of $W_\mathrm{p}$ are strongly limited by short duration time of acceleration also known as age-limited acceleration \citep[cf.][]{2010A&A...513A..17O}. On the other hand, other SNRs with an older age more than $\sim$8000 yrs show a steady decrease of $W_\mathrm{p}$ as SNRs get older. This trend could be understood considering the energy dependent diffusion of CRs \citep[e.g.,][]{1996A&A...309..917A,2007Ap&SS.309..365G}. In other words, in-situ values of $W_\mathrm{p}$ have been decreased due to CR escape from the SNR. In fact, hadron-dominant gamma-rays have been detected in nearby giant molecular clouds of W44, suggesting that the molecular clouds are illuminated by CR protons escaped from W44 \citep[e.g.,][]{2012ApJ...749L..35U,2020ApJ...896L..23P}. The authors suggested an actual value of $W_\mathrm{p}$ including escaped CRs is $\sim$$10^{50}$ erg, corresponding to 10\% of the typical kinematic energy released by a supernova explosion. In any case, W49B shows one of the highest in-situ values of $W_\mathrm{p}$ in the gamma-ray bright SNRs, which imply that the escape (diffusion) of CRs is not significant at the moment. Further gamma-ray observations using the Cherenkov Telescope Array (CTA) will unveil a transition phase from the age-limited acceleration to escape dominant stage in detail.

\section{CONCLUSIONS}\label{sec:conclusions}
We summarize the primary conclusions as follows:
\begin{enumerate}
\item New ALMA ACA CO($J$ = 2--1) observations at $\sim$$7''$ resolution have revealed the spatial and kinematic distributions of three candidates of interacting molecular clouds with the mixed-morphorogy SNR W49B, velocities of which are $\sim$10 km s$^{-1}$, $\sim$40 km s$^{-1}$, and $\sim$60 km s$^{-1}$. We found that western molecular clouds at $\sim$10 km s$^{-1}$ are obviously along with both the radio continuum boundary and inside filaments as well as the deformed X-ray shell, suggesting that shock-cloud interactions occurred. The 10 km s$^{-1}$ cloud also shows higher kinetic temperature of $\sim$20--{60} K than the reference clouds at {15} K, indicating that modest shock heating also occurred. The presence of a wind-bubble with an expanding velocity of $\sim$6 km s$^{-1}$ provides further evidence for the association of the 10 km s$^{-1}$ cloud.
\item The barrel-like structure of Fe-rich ejecta is mainly located on the void of dense molecular clouds, where a cool plasma component is enhanced. We propose a possible scenario that the barrel-like structure of Fe-rich ejecta was formed not only by the asymmetric supernova explosion, but also by interactions with dense molecular clouds. A supernova explosion occurred within the cylinder-like gaseous medium and then Fe-rich ejecta was efficiently heated-up at higher densities by multiple-reflected shocks formed by shock-cloud interactions. 
\item The electron temperature $kT_\mathrm{e}$ of recombining plasma from Fe He$\alpha$ shows a negative correlation with the peak integrated intensity of CO line emission in the 10 km s$^{-1}$ cloud. More precisely, $kT_\mathrm{e}$ values in high-temperature recombining plasma are increasing from the west (cloud rich) regions to the east (cloud poor) regions, suggesting the thermal conduction origin. Note that this finding does not rule out the adiabatic cooling scenario in the low-temperature recombining plasma in W49B which was previously discussed \citep[][]{2020ApJ...893...90S,2020ApJ...903..108H}.
\item The total energy of CR protons $W_\mathrm{p}$ is estimated to be $\sim$$2 \times 10^{49}$ erg, which is one of the highest values in gamma-ray bright SNRs. We found that in-situ values of $W_\mathrm{p}$ in gamma-ray SNRs increase with age for the young group (with the age less than $\sim$6000 yr). On the other hand, other older SNRs show a steady decrease of $W_\mathrm{p}$ as SNRs get older due to the escapes/diffusion effect of CRs. We frame a hypothesis that W49B is undergoing an age-limited acceleration without a significant escape or diffusion of CRs from the SNR.
\end{enumerate}

\section*{Acknowledgements}
We are grateful Hiroya Yamaguchi for providing us with the {\it{NuSTAR}} data points used in this paper. This paper makes use of the following ALMA data: ADS/JAO.ALMA\#2018.1.01780.S. ALMA is a partnership of ESO (representing its member states), NSF (USA) and NINS (Japan), together with NRC (Canada), MOST and ASIAA (Taiwan), and KASI (Republic of Korea), in cooperation with the Republic of Chile. The Joint ALMA Observatory is operated by ESO, AUI/NRAO and NAOJ. The scientific results reported in this article are based on data obtained from the {\it{Chandra}} Data Archive (Obs IDs: 117, 13440, and 13441). This research has made use of software provided by the {\it{Chandra}} X-ray Center (CXC) in the application packages CIAO (v 4.12). This work was supported by JSPS KAKENHI Grant Numbers JP19H05075 (H. Sano), and JP21H01136 (H. Sano). K. Tokuda was supported by NAOJ ALMA Scientific Research Grant Number of 2016-03B. We are also grateful to the anonymous referee for useful comments which helped the authors to improve the paper significantly.

\software{CASA \citep[v 5.6.0.:][]{2007ASPC..376..127M}, CIAO \citep[v 4.12:][]{2006SPIE.6270E..1VF}, CALDB \citep[v 4.9.1][]{2007ChNew..14...33G}.}

\facilities{ALMA, Chandra, NuSTAR, Very Large Array (VLA), Nobeyama 45-m Telescope, James Clerk Maxwell Telescope (JCMT).}

\end{document}